\documentclass[12pt, draftclsnofoot, onecolumn,comsoc]{IEEEtran}

\usepackage{cite}
\usepackage[dvips]{graphicx}
\usepackage[cmex10]{amsmath}
\usepackage{caption}
\usepackage{amsfonts}
\usepackage{amssymb}
\usepackage{float}
\usepackage{array}
\usepackage{rotating}

\newcolumntype{L}[1]{>{\raggedright\let\newline\\\arraybackslash\hspace{0pt}}m{#1}}
\newcolumntype{C}[1]{>{\centering\let\newline\\\arraybackslash\hspace{0pt}}m{#1}}
\newcolumntype{R}[1]{>{\raggedleft\let\newline\\\arraybackslash\hspace{0pt}}m{#1}}

\usepackage{subfig}
\usepackage{url}

\usepackage{xcolor}
\usepackage[utf8]{inputenc}
\usepackage{amsthm}
\usepackage{thmtools}

\newtheoremstyle{mytheoremstyle} % name
{\topsep}                    % Space above
{0pt}                    % Space below
{\itshape}                   % Body font
{}                           % Indent amount
{\bfseries}                   % Theorem head font
{.}                          % Punctuation after theorem head
{.5em}                       % Space after theorem head
{}  % Theorem head spec (can be left empty, meaning ‘normal’)

\theoremstyle{mytheoremstyle}

\newtheorem{theorem}{Theorem}
\newtheorem{lemma}{Lemma}
\newtheorem{corollary}{Corollary}
\newtheorem{remark}{Remark}

\def\h{\mathbf{h}}
\def\n{\mathbf{n}}

\def\s{\mathbf{s}}
\def\v{\mathbf{v}}

\def\x{\mathbf{x}}
\def\y{\mathbf{y}}
\def\z{\mathbf{z}}

\def\I{\mathbf{I}}
\def\N{\mathbf{N}}

\def\X{\mathbf{X}}

\def\Z{\mathbf{Z}}

\def\rmx{\mathrm{x}}

\def\bphi{\text{\boldmath$\phi$}}
\def\bvphi{\text{\boldmath$\varphi$}}

\def\rchi{\text{\raisebox{2pt}{$\chi$}}}

\def\E{\mathbb{E}}
\def\C{\mathbb{C}}
\def\R{\mathbb{R}}

\def\CN{\mathcal{CN}}
\def\cA{\mathcal{A}}

\def\cC{\mathcal{C}}
\def\cD{\mathcal{D}}

\def\sB{\mathsf{B}}
\def\sG{\mathsf{G}}
\def\sJ{\mathsf{J}}
\def\sT{\mathsf{T}}

\def\Bw{B_{\textsc{w}}}

\def\SINR{\mathrm{SINR}} 
\def\SIR{\mathrm{SIR}}

\def\Var{\mathrm{Var}} 
 
\def\BS{\mathrm{BS}} 
\def\UE{\mathrm{UE}} 
\def\AchR{\mathrm{R}} 
\def\EE{\mathrm{EE}} 
\def \Ptx{\mathrm{P}_\textsc{tx}} 

\def \Ptx{\mathrm{P}_\textsc{tx}}
\def \Plp{\mathrm{P}_\textsc{lp}}
\def \Pce{\mathrm{P}_\textsc{ce}}
\def \Pbs{\mathrm{P}_\textsc{bs}}

\newcommand{\bb}[1]{\mathbb{#1}}

\renewcommand{\rm}[1]{\mathrm{#1}}
\renewcommand{\cal}[1]{\mathcal{#1}}

\usepackage{tikz}
\usepackage{pgfplots}
\pgfplotsset{
	tick label style={font=\small},
	label style={font=\small},
	legend style={font=\small},
}

\usepackage{transparent}

 \usepackage{import}

\IEEEoverridecommandlockouts
\allowdisplaybreaks
\begin{document}

\title{Spectral and Energy Efficiency of Superimposed Pilots in Uplink Massive MIMO}

\author{Daniel~Verenzuela,~\IEEEmembership{Student~Member,~IEEE,}
        Emil~Bj\"{o}rnson,~\IEEEmembership{Member,~IEEE,}
        Luca~Sanguinetti,~\IEEEmembership{Senior~Member,~IEEE.}
\thanks{D. Verenzuela and E. Bj\"{o}rnson are with the Department
of Electrical Engineering (ISY), Link\"{o}ping University, Link\"{o}ping, SE-58183 Sweden (e-mail: daniel.verenzuela@liu.se; emil.bjornson@liu.se).}% <-this % 
\thanks{L. Sanguinetti is with the Dipartimento di Ingegneria dell'Informazione, University of Pisa, Pisa, Italy, and with the Large Networks and System Group (LANEAS), CentraleSup\'elec, Universit\'e Paris-Saclay, Gif-sur-Yvette, France (e-mail: luca.sanguinetti@unipi.it).}
\thanks{This paper has received funding from ELLIIT and the Swedish Foundation for Strategic Research (SFF). The work of L. Sanguinetti was supported in part by ERC Starting MORE under Grant 305123.}
\thanks{\color{black} A preliminary version \cite{DV_EB_LS_SE_SP_conf}  of this work will be presented at IEEE GLOBECOM 2017. }}

\maketitle
%\vspace*{-25pt}
\begin{abstract}
Next generation wireless networks aim at providing substantial improvements in spectral efficiency (SE) and energy efficiency (EE). Massive MIMO has been proved to be a viable technology to achieve these goals by spatially multiplexing several users using many base station (BS) antennas. A potential limitation of Massive MIMO in multicell systems is pilot contamination, which arises in the channel estimation process from the interference caused by reusing pilots in neighboring cells. A standard method to reduce pilot contamination, known as regular pilot (RP), is to adjust the length of pilot sequences while transmitting data and pilot symbols disjointly. An alternative method, called superimposed pilot (SP), sends a superposition of pilot and data symbols. This allows to use longer pilots which, in turn, reduces pilot contamination. We consider the uplink of a multicell Massive MIMO network using maximum ratio combining detection and compare RP and SP in terms of SE and EE. To this end, we derive rigorous closed-form achievable rates with SP under a practical random BS deployment. We prove that the reduction of pilot contamination with SP is outweighed by the additional coherent and non-coherent interference. Numerical results show that when both methods are optimized, RP achieves comparable SE and EE to SP in practical scenarios.
\end{abstract}

\IEEEpeerreviewmaketitle
\section{Introduction}

The development of cellular networks is lead by the continuous increase in mobile data traffic \cite{cisco_GMDT_2014}. The design of future cellular networks aims at handling 1000$\times$ more data traffic per unit area \cite{METIS_D8_4}. Meanwhile, the energy consumption of mobile communication systems is of great economical and ecological concerns \cite{Co2_footprint}. Massive multiple-input multiple-output (MIMO) is considered as one of the most promising technology to jointly improve spectral efficiency (SE) and energy efficiency (EE) \cite{ev_wireless_comm, Hien_bounds,Emil_power_model,Emil_MAMIMO_small_cells,T_marzetta_total_EE}. The key idea of Massive MIMO is to utilize a large number of antennas (e.g., hundreds or thousands) at the base stations (BSs) to communicate coherently with several (e.g., tens or hundreds) user equipments (UEs) by virtue of spatial multiplexing \cite{marzetta_MAMIMO,T_Marzetta_Ma_MIMO_book}. 

The acquisition of channel state information (CSI) at the BS is essential in Massive MIMO. A time division duplexing (TDD) system is usually proposed to avoid the large overhead of downlink channel training and feedback \cite{T_Marzetta_Ma_MIMO_book}. Uplink pilot sequences are transmitted by the UEs and channel reciprocity is exploited at the BS to coherently detect data from UEs in the uplink and precode data in the downlink. The time and frequency interval, over which the channel can be considered to remain static and frequency flat, called the coherence block, has a limited size and, in turn, there is a finite number of orthogonal pilot sequences that are available for channel estimation. Therefore, in multicell systems the pilot sequences need to be reused across cells. This creates coherent interference, called pilot contamination, between UEs that share the same pilots, which reduces the quality of channel estimates and affects the SE. The pilot contamination has been widely investigated in the literature. In \cite{MAMI_low_nr_ant, coord_ch_est, Bjornson_pilot_cont_not_lim_jrnl, Julia_pilot_decont, Hien_EVD_pilot, blind_pilot_decont}, the same set of pilot sequences is assumed to be reused in all the cells and pilot contamination is mitigated by exploiting spatial channel correlation \cite{MAMI_low_nr_ant, coord_ch_est,Bjornson_pilot_cont_not_lim_jrnl} or data covariance matrices \cite{Julia_pilot_decont, Hien_EVD_pilot, blind_pilot_decont}. Another approach is to have longer pilot sequences than the number of served UEs per cell to reduce the number of cells utilizing the same pilot \cite{T_marzetta_total_EE,T_marzetta_non_asym, Emil_pilot_SE, Emil_pilot_cluster}. This method can effectively reduce pilot contamination at the cost of an increased estimation overhead that, in turn, decreases the amount of data symbols transmitted per coherence block. This tradeoff is studied in \cite{Emil_pilot_SE} under a hexagonal cell deployment and it turned out that a fraction between 5\% and 40\% of the coherence block should be used for pilots. 

In all the aforementioned works, the transmission of pilot and data symbols is done separately within the coherence block to reduce interference in the channel estimation process. This method is known in the literature as regular pilot (RP) transmission. In \cite{ Hoeher99_ch_est_SP, GTZhou03_SP_ch_est_1ord_st, Dong_SiP_SigProcess_04, SP_stat_fading_MIMO_2017, SIP_part1_KU_SA, SIP_part2_KU_SA, VT_SP_approx_2016}, the authors explore an alternative method that relies on the simultaneous transmission of pilot and data signals. This method is referred to as superimposed pilot (SP) and allows to increase the amount of samples that can be used for channel estimation and data transmission. By using SP, \cite{Hoeher99_ch_est_SP} propose an optimal coherent receiver based on the Viterbi algorithm. Linear channel estimation methods of finite impulse response channels for single-input single-output (SISO) systems are considered in \cite{GTZhou03_SP_ch_est_1ord_st} with only knowledge of the first order statistics. In \cite{Dong_SiP_SigProcess_04}, the authors compare SP and RP under Gauss-Markov flat fading SISO channels under a practical setup where channels change rapidly and UEs have low signal-to-noise ratios (SNRs). The results show that SP provides better performance than RP in terms of uncoded bit-error-rate (BER) and mean squared error (MSE) of channel estimates. {\color{black}Similar results have been found for stationary MIMO fading channels in \cite{SP_stat_fading_MIMO_2017}. In the aforementioned works \cite{ Hoeher99_ch_est_SP, GTZhou03_SP_ch_est_1ord_st, Dong_SiP_SigProcess_04, SP_stat_fading_MIMO_2017}, the authors focus on a single cell or single user scenario.  Recently, \cite{SIP_part1_KU_SA, SIP_part2_KU_SA, VT_SP_approx_2016} have shown that SP achieves promising results in multicell Massive MIMO systems.} In particular, UEs transmit a linear combination of pilot and data symbols within the whole coherence block. This allows the use of longer pilot sequences, which can thus be reused less frequently in the network. This allows to reduce pilot contamination, which could, in principle, improve the SE. However, sending pilot and data signals simultaneously causes interference in the channel estimation process from data symbols. This degrades the estimation quality and creates correlation between channel estimates and data. Moreover, the use of longer pilots increases the computational complexity of channel estimation and data detection. This, in turn, consumes more power and may eventually reduce the EE of the network. In summary, the use of SP in Massive MIMO systems introduces new sources of interference and increases the consumed power. All this may limit the practical gains of SP methods in terms of SE and EE. 

The aim of this paper is to evaluate the performance of SP in the uplink of a multicell Massive MIMO system and make comparisons with RP.
To this end, we derive rigorous closed-form rate expressions with SP when using maximum ratio combining (MRC). {\color{black}This stands in contrast to prior works, \cite{SIP_part1_KU_SA, VT_SP_approx_2016}, which deal with approximate expressions of signal-to-interference-plus-noise ratios (SINRs) and mean square errors (MSEs). The analysis provided in this paper holds true for any number of BS antennas (not just for a large number).} These formulas provide valuable insights into identifying all the interference sources, their impact on the SE and their relationship with the other system parameters. The provided expressions are then used to perform the asymptotic analysis (corresponding to the large number of BS antennas regime) of the network, which allows to identify the conditions under which either RP or SP provide greater rates. Then, in order to properly study the effect associated with intercell interference in a large practical network with an irregular BS deployment, we adopt the stochastic geometry framework developed in \cite{Emil_MAMIMO_small_cells} wherein BSs are spatially distributed according to a homogeneous Poisson point process (PPP). Within this setting, we calculate closed-form lower bounds of the achievable rates averaged over the UEs' spatial distribution. This provides powerful insights into the interplay of system parameters without requiring the use of heavy numerical simulations. Such lower bounds are then used to compute the EE of the network with both RP and SP taking into account the power consumed by transmission and circuitry. Numerical results are used to show that, when both methods are optimized, RP provides comparable SE and EE to SP in practical scenarios. 

The remainder of this paper is organized as follows. Section~\ref{sec:sys_mdel} introduces the network model. In Section~\ref{sec:ch_est}, the channel estimation process with RP and SP is detailed whereas the achievable rates with MRC are computed in Section~\ref{sec:ach_rates}. Section~\ref{sec:ach_rate_analysis} presents detailed analytical comparisons between RP and SP. In Section~\ref{sec:sto_geometry}, the average achievable rates are first computed for a random network deployment (based on stochastic geometry) and then used for computing the EE. Section~\ref{sec:num_result} illustrates numerical results while Section~\ref{sec:conclusion} concludes our work.

\subsubsection*{Notation}
We denote vectors by lower-case bold-face letters (e.g., $\x$)\footnote{$[\x]_j$ refers to the $j^{th}$ element of $\x$.} and matrices by bold-face capital letters (e.g., $\X$).\footnote{$[\X]_{j}$ denotes the $j^{th}$ column of $\X$ and $[\X]_{ij}$ refers to the $i^{th}$ row and $j^{th}$ column element of $\X$.} The operators $\E\{\cdot\}$ and $\E\{\cdot|y\}$ represent expected value  and  
 expected value conditioned on a realization of the random variable $y$,\footnote{We abuse the notation in conditional expectations by referring to the random variable and its realization with the same letter.} respectively. The notation $|\cdot|$ represents the absolute value and $\|\cdot\|$ denotes the Euclidean norm. We denote the transpose, conjugate transpose and conjugate operators as $(\cdot)^T$, $(\cdot)^H$ and $(\cdot)^*$, respectively. We denote by $\I_M$ the identity matrix of size $M\times M$ and $\CN(\cdot,\cdot)$ indicates a circularly symmetric complex Gaussian distribution. To denote the set of real and complex numbers we use $\R$ and $\C$, respectively, while $\Re(\cdot)$ is the real part. $\Gamma(\cdot)$ denotes the Gamma function.
\section{Network Model}
\label{sec:sys_mdel}
We consider the uplink of a multicell Massive MIMO network where each BS has $M$ antennas and serves $K$ single-antenna UEs. We define $\Phi_{D}$ the set containing all BSs, where $D$ denotes the density of BSs per unit area (measured in BS/km$^2$). Note that this definition does not require the BSs to be distributed in any specific manner. However, a stochastic geometry framework will be used later on in Section~\ref{sec:sto_geometry} to model the BS distribution. Without loss of generality, the following analysis is focused on an arbitrary BS, denoted as $\BS_0$ serving UEs in cell $0$, and an arbitrary UE $k$ in cell $0$, denoted as $\UE_{0k}$. We define $\Psi_{D} = \Phi_{D}\backslash\{0\}$ as the set containing all other BSs than $\BS_0$.

\begin{figure}[!t] %
	\centering %
\def\svgwidth{0.55\textwidth}	
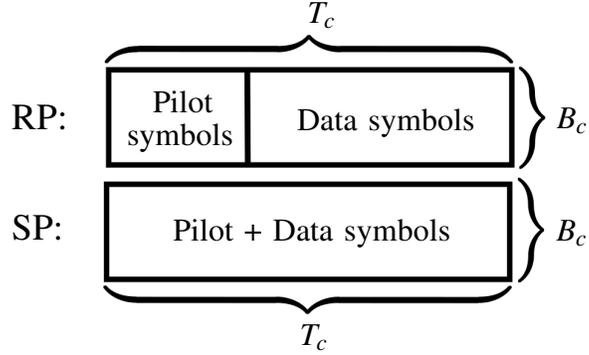
%\\[-15pt]
	\caption{ Transmission protocol with RP and SP methods.} 
	\label{fig:Tx_protocol} 
\end{figure}

We consider a network with bandwidth $\Bw$. The communication channels are modeled as block fading where each channel is considered to be constant over a coherence block of time duration $T_c$ and bandwidth $B_c$.\footnote{{\color{black} In an OFDM system, the coherence bandwidth $B_c$ includes several subcarriers---see \cite{T_Marzetta_Ma_MIMO_book} for more details.}} The total bandwidth is equally divided among all coherence blocks, which means that $\Bw/B_c$ is an integer number, and each block contains $\tau_c = B_c T_c$ complex samples. {\color{black} We assume uncorrelated Rayleigh fading channels since this is the first rigorous capacity analysis with SP in a multicell scenario. As done with RP, we believe that it is helpful to first develop fundamental theory for uncorrelated channels and then to extend it to correlated channels. Therefore, this is left for future work. Moreover, since uncorrelated fading corresponds to the worst-case scenario for pilot contamination and SP aims at mitigating this effect, this analysis gives insights into the main benefits of SP. In addition, the achievable rates under uncorrelated Rayleigh fading are close to those under practical measured channels with non-line-of-sight and spatially distributed UEs \cite{FTufvesson15_measured_MAMIMO}.} We denote by ${\h_{ll'i}\in \C^{M}}$ the channel between the $M$ antennas of $\BS_l$ and $\UE_{l'i}$ in which the small-scale fading (SSF) is modeled as $\h_{ll'i} \sim \CN\left( \mathbf{0}, \beta_{ll'i}\I_M\right)$ $\forall l,l'\in\Phi_{D}$ and $i \in \{1,\ldots,K\}$ with ${\beta_{ll'i}\geq 0}$ being the large-scale fading (LSF) coefficient between $\BS_l$ and $\UE_{l'i}$. We assume that the distance between UEs and BSs is large enough to consider $\beta_{ll'i}$ to be the same for all BS antennas. The received signal $\y_0\in \C^{M}$ at $\BS_0$ is 
\begin{align}
\y_0 = \sum_{l'\in \Phi_{D}}\sum_{i=1}^{K} \h_{0l'i} \rmx_{l'i} + \n_0 
\label{eq:gen_rec_signal}
\end{align}
where $\n_0\in \C^{M}$ is the noise vector distributed as $\n_0\sim\CN(\mathbf{0},\sigma^2\I_M)$ and $\rmx_{l'i}$ represents the transmitted signal from $\UE_{l'i}$ in one arbitrary sample of the coherence block. The transmitted signal can be used for data, pilots or a superposition of the two depending on the employed method. We analyze the two transmission methods illustrated in Fig.~\ref{fig:Tx_protocol}: RP, called time-multiplexed in \cite{SIP_part1_KU_SA},  and SP. With RP, data and pilot symbols are transmitted separately in each coherence block. Therefore, $\rmx_{l'i}$ contains only one of the two in each sample of the coherence block. With SP, pilot and data symbols are transmitted simultaneously during the whole coherence block and thus $\rmx_{l'i}$ contains a superposition of the two in each sample.

\section{Channel Estimation}
\label{sec:ch_est}
To estimate the channels, we use standard linear minimum mean squared error (LMMSE) techniques \cite{steve_M_kay} with both RP and SP.
\subsection{Regular pilots}
\label{subsec:ch_est_RP}
We consider a transmission protocol where $\tau_p$ out of the $\tau_c$ samples in each coherence block are reserved for pilot sequences, which leaves a fraction $1 - \tau_p/\tau_c$ of samples for data transmission. {\color{black}We consider a set of $\tau_p$ orthogonal pilot sequences of length $\tau_p$. Each BS allocates $K\leq \tau_p$ different pilot sequences to the UEs served in its cell. We denote as ${\bphi_{l'i} \in \C^{\tau_p},\;\forall l' \in \Phi_{D}, \; i\in \{1,\ldots,K\} }$ the pilot sequence assigned to $\UE_{l'i}$ with ${| [\bphi_{l'i} ]_j | = 1}$, ${\forall j \in \{1,\ldots,\tau_p\}}$. To identify the UEs in different cells that share the same pilot as $\UE_{0k}$ (including $\UE_{0k}$), we define the set $\cal{P}_{0k}^\textsc{rp} = \left\{ \{l',i\} \text{ : } \bphi_{0k}^H\bphi_{l'i} \neq 0 \right\}$.} $\UE_{0k}$ transmits its pilot sequence $\bphi_{0k}^T$ along with all other UEs in the network over $\tau_p$ instances of \eqref{eq:gen_rec_signal}. At $\BS_0$, this yields the received signal $\Z_{0k}^\textsc{rp}\in \C^{M\times \tau_p}$ given~by
\begin{equation}
\begin{aligned}
\Z_{0k}^\textsc{rp}  &=   \sum_{l'\in\Phi_{D}}\sum_{i=1}^{K} \sqrt{q_{l'i}} \h_{0l'i}\bphi_{l'i}^T + \bar{\N}_0
\label{eq:rec_sig_pilot}
\end{aligned}
\end{equation}
where $q_{l'i}$ is the transmission power of the pilot symbols from $\UE_{l'i}$ and $\bar{\N}_0$ is the noise matrix with i.i.d. elements distributed as $[\bar{\N}_0]_{mj} \sim\CN \left(0, \sigma^2\right) \forall m \in \{1,\ldots,M\},\;{j \in \{1,\ldots,\tau_p\}} $ with $\sigma^2$ being the noise variance. By multiplying $\Z_{0k}^\textsc{rp}$ with $\bphi_{0k}^*/\sqrt{\tau_p}$, the received pilot signal is correlated with the pilot sequence corresponding to $\UE_{0k}$, which is equivalent to despreading the received signal. This operation yields $\z_{0k}^\textsc{rp}\in \C^{M}$ given by
{\color{black}\begin{align}\label{z{0k}}
\z_{0k}^\textsc{rp} = \Z_{0k}^\textsc{rp}\frac{\bphi_{0k}^*}{\sqrt{\tau_p}}
&=  \sum_{\{l',i\}\in\cal{P}_{0k}^\textsc{rp}} \sqrt{q_{l'i}\tau_p} \h_{0l'i} +  \bar{\n}_0
\end{align}}%
where $\bar{\n}_0 = \bar{\N}_0\bphi_{0k}^*/\sqrt{\tau_p}$ is a noise vector distributed as ${\bar{\n}_0\sim\CN\left(\mathbf{0},\sigma^2\I_M\right) }$. Notice that no useful information is lost in the despreading operation, given that any signal in the orthogonal complement of $\bphi_{0k}$ is independent of $\z_{0k}^\textsc{rp}$. Therefore, $\z_{0k}^\textsc{rp}$ in \eqref{z{0k}} is a sufficient statistic for estimating the channel $\h_{00k}$ between $\BS_0$ and $\UE_{0k}$. The minimum mean squared error (MMSE) estimate of $\h_{00k}$ is given by the next lemma.
\begin{lemma}
	With RP, the MMSE estimate of $\h_{00k}$ is
	\begin{align}
	\hat{\h}_{00k} = \frac{\bar{\gamma}_{0k}^\textsc{rp}}{\sqrt{ q_{0k} \tau_p}} \z_{0k}^\textsc{rp}
	\end{align}
	with
	{\color{black}\begin{align}
	&\bar{\gamma}_{0k}^\textsc{rp} = \frac{q_{0k}\tau_p \beta_{00k}}{ \sum_{\{l',i\}\in\cal{P}_{0k}^\textsc{rp}} q_{l'i}\tau_p \beta_{0l'i}  + \sigma^2}
	\end{align}}
	and has covariance matrix given by
	{\color{black}\begin{align}
	&\E\left\{\hat{\h}_{00k} \hat{\h}_{00k}^H \right\}= \beta_{00k}\bar{\gamma}_{0k}^\textsc{rp}\I_M.
	\label{eq:est_quality_rp}
	\end{align}}
\end{lemma}
\begin{IEEEproof}
	It follows from applying standard LMMSE techniques\mbox{\cite[Ch.~12]{steve_M_kay}} to the problem at hand. Since $\z_{0k}^\textsc{rp}$ contains a Gaussian unknown signal plus independent Gaussian interference and noise, the LMMSE estimator coincides with the true MMSE estimator.
\end{IEEEproof}
The parameter $\bar{\gamma}_{0k}^\textsc{rp} \in [0,1]$ indicates the quality of channel estimates. Notice that, as the length $\tau_p$ of the pilot sequences increases, $\bar{\gamma}_{0k}^\textsc{rp}$ also increases since the noise term becomes less significant {\color{black} and the cardinality of $\cal{P}_{0k}^\textsc{rp}$} decreases with $\tau_p$. This means that, as $\tau_p$ increases, the variance of the channel estimates approaches the variance of the true channels and estimation errors vanish. However, in practical applications $\tau_p\leq \tau_c$. Since $\tau_c$ is limited by the physical properties of the channel, there will always be an estimation error due to pilot contamination and noise. The key point to notice is that for scenarios where $\tau_c$ is much larger than $K$, the channel estimates with RP can be improved by letting $\tau_p$ be larger than $K$. 

\subsection{Superimposed pilots}
\label{subsec:ch_est_SP}
With SP, all the $\tau_c$ samples of the coherence block are used for transmitting pilot and data symbols. {\color{black} We consider $\tau_c$ orthogonal pilot sequences of length $\tau_c$ samples. Each BS selects $K\leq \tau_c$ different pilots and assigns them to its UEs. We denote as ${\bvphi_{l'i} \in \C^{\tau_c}}$, ${\forall l' \in \Phi_{D}}$, ${i\in \{1,\ldots,K\} }$ the pilot sequence assigned to $\UE_{l'i}$ with ${| [\bvphi_{l'i} ]_j | = 1}$, ${\forall j \in \{1,\ldots,\tau_c\}}$.\footnote{{\color{black} Note that since the modulus of each pilot symbol is one, the peak-to-average power ratio of the transmitted SP signal does not increase when adding the pilot symbols.}} The set $\cal{P}_{0k}^{\textsc{sp}} = \left\{ \{l',i\}\text{ : } \bvphi_{0k}^H\bvphi_{l'i} \neq 0 \right\}$ contains the indices of the UEs using the same pilot as $\UE_{0k}$ (including  $\UE_{0k}$).} $\UE_{0k}$ transmits a superposition of the pilot sequence $\bvphi_{0k}^T$ and the data signal $\s_{0k}^T$ along with all other UEs in the network over $\tau_c$ instances of \eqref{eq:gen_rec_signal}. This yields an $M\times \tau_c$ received signal at $\BS_0$ given by
\begin{equation}
\label{eq:rec_sig_pilot_SIP}
\Z_{0k}^\textsc{sp}  = \sum_{l'\in\Phi_{D}}\sum_{i=1}^{K} \sqrt{q_{l'i}} \h_{0l'i}\bvphi_{l'i}^T 
%\\&\mkern 20mu
+ \sum_{l'\in \Phi_{D}} \sum_{i=1}^{K} \sqrt{p_{l'i}} \h_{0l'i}\s_{l'i}^T + \N_0% \in \C^{M\times \tau_c}
\end{equation}
where $p_{l'i}$ and $q_{l'i}$ are the transmission powers of the data and pilot symbols, respectively,  transmitted by $\UE_{l'i}$. The vector $\s_{l'i} \in \C^{\tau_c}$ contains the data symbols transmitted in the whole coherence block. We assume the data symbols to be i.i.d. as $\s_{l'i}\sim\CN\left(\mathbf{0},\I_{\tau_c}\right)$. The noise matrix is defined as ${\N_0 = \left[\n_{01}, \ldots, \n_{0\tau_c} \right]}$ with i.i.d. columns distributed as $\n_{0j} \sim \CN \left(0, \sigma^2\I_M\right)$ $\forall j \in \{1,\ldots,\tau_c\}$. By multiplying $\Z_{0k}^\textsc{sp}$ with $\bvphi_{0k}^*/\sqrt{\tau_c}$, we obtain
{\color{black}\begin{equation}% 
\begin{aligned}
\z_{0k}^\textsc{sp}\! = \!\Z_{0k}^\textsc{sp}\frac{\bvphi_{0k}^*}{\sqrt{\tau_c}}
&\!=\!  \sum_{\{l',i\}\in\cal{P}_{0k}^\textsc{sp}} \sqrt{q_{l'i}\tau_c} \h_{0l'i}  
\!+\! \sum_{l'\in \Phi_{D}} \!\sum_{i=1}^{K} \!\sqrt{\frac{p_{l'i}}{\tau_c}} \h_{0l'i}\s_{l'i}^T\bvphi_{0k}^* \!+\!  \sum_{j=1}^{\tau_c} \n_{0j}\frac{\left[\bvphi_{0k}\right]_{j}^*}{\sqrt{\tau_c}}
\label{eq:z_0k_ch_est_rec_sig}
\end{aligned}
\end{equation}}
which is then used to compute the LMMSE estimate the channel between $\BS_0$ and $\UE_{0k}$. 
\begin{lemma}
	With SP, the LMMSE estimate of the channel $\h_{00k}$ is
	\begin{align}
	\hat{\h}_{00k} = \frac{\bar{\gamma}_{0k}^\textsc{sp}}{\sqrt{ q_{0k} \tau_c}} \z_{0k}
	\end{align}
	where
	{\color{black}\begin{align}
	&\bar{\gamma}_{0k}^\textsc{sp} = \frac{q_{0k}\tau_c \beta_{00k}}{\sum_{\{l',i\}\in\cal{P}_{0k}^\textsc{sp}} q_{l'i} \tau_c\beta_{0l'i}  +\sum_{l'\in\Phi_{D}} \sum_{i =1}^K p_{l'i}\beta_{0l'i} + \sigma^2}\:.
	\label{eq:ch_quality_SP}
	\end{align}}
	The covariance matrix of $\hat{\h}_{00k}$ is 
	{\color{black}\begin{align}
	&\E\left\{\hat{\h}_{00k} \hat{\h}_{00k}^H \right\}= \bar{\gamma}_{0k}^\textsc{sp}\beta_{00k}\I_M.
	\label{eq:est_quality_sip}
	\end{align}}
	\label{lem:ch_est_SIP}
\end{lemma}
\begin{IEEEproof}
	It follows from applying standard LMMSE estimation techniques  \cite[Ch. 12]{steve_M_kay} to the problem at hand.
\end{IEEEproof}
The parameter $\bar{\gamma}_{0k}^\textsc{sp} \in [0,1]$ indicates the quality of the channel estimates. From \eqref{eq:ch_quality_SP}, it follows that the interference caused by data symbols is $\tau_c$-times less influential than the pilot interference from UEs that use the same pilot as $\UE_{0k}$. Moreover, as the length $\tau_c$ of the pilot sequences increases, $\bar{\gamma}_{0k}^\textsc{sp}$ approaches one since the data interference and noise become less influential {\color{black} and the cardinality of $\cal{P}_{0k}^\textsc{sp}$ decreases with $\tau_c$}. This means that the variance of the channel estimates approaches the variance of the true channels. However, in practical applications $\tau_c$ is limited and thus there will always be an estimation error due to pilot contamination as well as interference from data signals and noise. 
{\color{black}\begin{remark}
		\label{rem:ch est_SP}
		The key difference between the channel estimates with RP and SP, apart from the number observations ($\tau_p$ with RP and $\tau_c$ with SP), is the presence of extra interference with SP due to the received data symbols (see the third term in the right-hand-side of \eqref{eq:z_0k_ch_est_rec_sig}). This interference not only reduces the quality of the channel estimates but it also:
		\begin{itemize}
			\item Changes the distribution of the channel estimates. The received signal $\z_{0k}^\textsc{sp}$  in \eqref{eq:z_0k_ch_est_rec_sig} is not Gaussian. Thus, the LMMSE estimate does not coincide with the true MMSE estimate and the channel estimates are only uncorrelated to the channel estimation errors but not independent, which stands in contrast to RP.
			\item Creates correlation between the channel estimates and received data symbols from all UEs. 	
		\end{itemize}
		These phenomena play a key role in the achievable rate analysis with SP and create extra interfering terms that cannot be obtained from the closed-form expressions provided in \cite{T_Marzetta_Ma_MIMO_book}.		
	\end{remark}}

\section{Achievable rates with MRC}
\label{sec:ach_rates}
To evaluate the performance of the network, we derive ergodic achievable rates by applying standard lower bounding techniques on the capacity (e.g., \cite{T_Marzetta_Ma_MIMO_book}). Since we consider a fixed bandwidth, the SE is obtained simply by scaling the achievable rates with $1/\Bw$. We assume that MRC is employed for data detection. Particularly, the estimates of the data symbols transmitted by $\UE_{0k}$ are obtained at $\BS_0$ by the inner product $\v_{00k}^H\y_0$ with $\v_{00k} = \upsilon_{00k} \hat{\h}_{00k}$, where ${\upsilon_{00k} = \frac{1}{\bar{\gamma}_{0k}^\textsc{rp}\sqrt{M \beta_{00k}}}}$ with RP and $\upsilon_{00k} = \frac{1}{\bar{\gamma}_{0k}^\textsc{sp}\sqrt{M \beta_{00k}}}$ with SP. These scaling factors are selected to provide an equivalent gain of $M \beta_{00k}$ for the desired signal with both methods. %{ 

{\color{black}To motivate the use of MRC, note that as $M\to \infty$, the directions of the channels $\h_{ll'i}/\|\h_{ll'i}\|$ of different UEs become asymptotically orthogonal. This is known as asymptotically favorable propagation. The squared norm of the channel scaled by $1/M$ converges to a deterministic number, which is known as channel hardening. When considering uncorrelated Rayleigh fading, these phenomena make the use of linear detection techniques like MRC asymptotically optimal as $M\to \infty$ \cite{T_Marzetta_Ma_MIMO_book}. In addition, the use of MRC has low complexity in the detection process and thereby low consumed power.}
{\color{black}
\subsection{Random Pilot allocation}
	\label{subsec:random_pilot_allc}
The key advantage that SP has with respect to RP is the ability to use the whole coherence block for both channel estimation and data detection. To obtain clear insights into the data rate performance with respect to the number of samples used of channel estimation, $\tau_p$ (with RP) and $\tau_c$ (with SP), we consider a random pilot allocation method with both RP and SP. In particular, we assume that each BS selects $K$, out of $\tau_p$ (with RP) or $\tau_c$ (with SP), distinct pilot sequences uniformly at random in each coherence block and allocates them to its served UEs. We define $\rchi_{l'i}^\textsc{rp} = \frac{\bphi_{0k}^H\bphi_{l'i}}{\tau_p} \in \{0,1\}$ and $\rchi_{l'i}^\textsc{sp} = \frac{\bvphi_{0k}^H\bvphi_{l'i}}{\tau_c}\in \{0,1\}$ as binary random variables to indicate if $\UE_{l'i}$ has the same pilot as $\UE_{0k}$ with RP and SP, respectively. Notice that BSs allocate pilots independently and that UEs within each cell have different pilots. This means that for $l'\neq 0$, $\sum_{i=1}^{K}\rchi_{l'i}^\textsc{rp}$ and $\sum_{i=1}^{K}\rchi_{l'i}^\textsc{sp}$ are Bernoulli distributed random variables with success probability $K/\tau_p$ and $K/\tau_c$, respectively. Thus, the following results hold:
\begin{align}
\label{eq:rnd_pilot_RP}
\bb{E} \left\{\sum\limits_{\{l',i\}\in\cal{P}_{0k}^\textsc{rp}\backslash\{0,k\}}^{K}\!\!\!\!\!\!q_{l'i} \beta_{0l'i} \right\} &\!=\!
\bb{E} \left\{\sum\limits_{l'\in\Psi_D }\sum\limits_{i=1}^{K}\!\rchi_{l'i}^{\textsc{rp}}q_{l'i} \beta_{0l'i} \right\}\!=\!
 \sum\limits_{l'\in\Psi_D }\frac{K}{\tau_p}\frac{1}{K}\left(\sum\limits_{i=1}^{K}\!q_{l'i} \beta_{0l'i} \right)\\
\bb{E} \left\{\sum\limits_{\{l',i\}\in\cal{P}_{0k}^\textsc{sp}\backslash\{0,k\}}^{K}\!\!\!\!\!\!q_{l'i} \beta_{0l'i} \right\} &\!=\!
\bb{E} \left\{\sum\limits_{l'\in\Psi_D }\sum\limits_{i=1}^{K}\!\rchi_{l'i}^{\textsc{sp}}q_{l'i} \beta_{0l'i} \right\}\!=\!
\sum\limits_{l'\in\Psi_D }\frac{K}{\tau_c}\frac{1}{K}\left(\sum\limits_{i=1}^{K}\!q_{l'i} \beta_{0l'i} \right)
\label{eq:rnd_pilot_SP}
\end{align}
which allow us to obtain achievable rate expressions that do not depend on the particular construction of the sets $\cal{P}_{0k}^\textsc{rp}$ and $\cal{P}_{0k}^\textsc{sp}$.
}

\subsection{Regular pilots}
\label{subsec:ach_rate_rp}
The received signal at $\BS_0$ with RP, for an arbitrary data symbol $j$ in the coherence block, is
\begin{equation}
\y_{0j}^\textsc{rp}  =  \sum_{i= 1}^K  \sqrt{p_{0i}} \h_{00i} [\s_{0i}]_j +  \sum_{l'\in\Psi_{D}}\sum_{i= 1}^K  \sqrt{p_{l'i}} \h_{0l'i} [\s_{l'i}]_j  + \n_{0j}%\in \C^{M }
\label{eq:rec_sig}
\end{equation}
where $\n_{0j}$ is the noise vector distributed as ${\n_{0j}\sim\CN\left(\mathbf{0},\sigma^2\I_M\right) }$.
To detect the data symbol from $\UE_{0k}$, the received signal $\y_{0j}^\textsc{rp}$ is combined with $\v_{00k}$ to obtain 
\begin{equation}
\begin{aligned}
\left[\hat{\s}_{0k} \right]_j &=\v_{00k}^H\y_{0j}^\textsc{rp} = \sqrt{p_{0k}}\E\left\{\v_{00k}^H\h_{00k} \right\} [\s_{0k}]_j + \sqrt{p_{0k}}\left(\v_{00k}^H\h_{00k} - \E\left\{\v_{00k}^H\h_{00k} \right\}\right) [\s_{0k}]_j\\
&\mkern 100mu +  \sum_{i\neq k}^K  \sqrt{p_{0i}} \v_{00k}^H\h_{00i} [\s_{0i}]_j +  \sum_{l'\in\Psi_{D}}\sum_{i= 1}^K  \sqrt{p_{l'i}} \v_{00k}^H\h_{0l'i} [\s_{l'i}]_j  + \v_{00k}^H\n_{0j} \;.
\label{eq:est_data_sig_rp}
\end{aligned}
\end{equation}
By treating the term $\sqrt{p_{0k}}\E\left\{\v_{00k}^H\h_{00k} \right\} [\s_{0k}]_j$ as the desired signal and the remaining ones in \eqref{eq:est_data_sig_rp} as effective noise, we have an equivalent SISO system with a deterministic channel and non-Gaussian effective noise, which is uncorrelated with the data symbol $[\s_{0k}]_j$. Moreover, the individual terms in the effective noise (second to last terms in \eqref{eq:est_data_sig_rp}) are also uncorrelated due to the fact that the data symbols from different UEs have zero mean and are independent among themselves and independent from the noise. In the next lemma, we provide an ergodic achievable rate, i.e., a lower bound on the capacity, of the system when using RP.   
{\color{black}\begin{lemma}
	\label{lem:ach_rate_ssf_RP}
	An ergodic achievable rate for $\UE_{0k}$ with RP and MRC detection is
	\begin{equation}
	\AchR_{0k}^\textsc{rp} = \Bw \left(1 - \frac{\tau_p}{\tau_c}\right)\log_2\left(1 + \SINR_{0k}^\textsc{rp}\right)
	\label{eq:rate_rp_gen}
	\end{equation}
	where $\SINR_{0k}^\textsc{rp}$ is the effective SINR of $\UE_{0k}$ given by
	\begin{align*}\allowdisplaybreaks
		\stepcounter{equation}\tag{\theequation}\label{eq:SINR_ssf_RP}
	\SINR_{0k}^\textsc{rp} &= \frac{p_{0k}\left|\E \!\left\{ \v_{00k}^H\h_{00k}\right\} \right|^2}{\sum\limits_{l'\in\Phi_{D}} \sum\limits_{i= 1}^K  \! p_{l'i} \E \!\left\{ \left|\v_{00k}^H\h_{0l'i}\right|^2\right\}- \left|\E \!\left\{ \v_{00k}^H\h_{00k}\right\} \right|^2 +\E\! \left\{ \left|\v_{00k}^H\n_0\right|^2 \right\} }\\			
	&=\frac{M p_{0k} \beta_{00k} }{ \frac{M}{\tau_p} \sum\limits_{l'\in\Psi_{D}}\sum\limits_{i=1}^{K} \frac{p_{l'i} q_{l'i}}{q_{0k}}\frac{\beta_{0l'i}^2}{\beta_{00k}}
		+   \frac{1}{\gamma_{0k}^\textsc{rp}}\left(\sum\limits_{l'\in\Phi_{D}} \sum\limits_{i=1}^{K}p_{l'i}\beta_{0l'i} + \sigma^2\right)}
		\stepcounter{equation}\tag{\theequation}
	\label{eq:SINR_rp_gen}
	\end{align*}	
	and 
	\begin{equation}
	\gamma_{0k}^\textsc{rp} = \E\left\{\frac{1}{\bar{\gamma}_{0k}^\textsc{rp}}\right\}^{-1} = \frac{q_{0k}\tau_p\beta_{00k}}{q_{0k} \tau_p \beta_{00k} + \sum\limits_{l'\in\Psi_{D}} \sum\limits_{i=1}^{K} q_{l'i}  \beta_{0l'i}+ \sigma^2 }.
	\label{eq:gamma_rp_gen}
	\end{equation}	
	The expectations in \eqref{eq:SINR_ssf_RP} are taken with respect to the SSF and the random pilot allocation. Note that the ergodic achievable rate with effective SINR given by \eqref{eq:SINR_ssf_RP} holds for any selection of $\v_{00k}$ and any channel distribution.
\end{lemma}}
\begin{IEEEproof}
	It follows from standard lower bounds \cite[Ch. 2]{T_Marzetta_Ma_MIMO_book} on the capacity between the transmitter and receiver when only knowledge of the average effective channel $\E\left\{\v_{00k}^H\h_{00k} \right\}$ is used to obtain an equivalent SISO system with a deterministic channel and non-Gaussian effective noise. The closed-form expression of the SINR follows the same approach as in \cite{Emil_pilot_cluster,Emil_MAMIMO_small_cells},\mbox{\cite[Ch. 4]{T_Marzetta_Ma_MIMO_book}} where the independence between the channel estimates and errors is used to compute the expectations in \eqref{eq:SINR_ssf_RP} in closed-form. {\color{black}In addition, the result in \eqref{eq:rnd_pilot_RP} is used to calculate the expectations with respect to $\rchi_{l'i}^\textsc{rp}$.}
\end{IEEEproof}
{\color{black}To mitigate the effect of pilot contamination with RP, we can increase the pilot overhead by selecting $\tau_p > K$. This improves the quality of channel estimates (see Section~\ref{subsec:ch_est_RP}) and reduces the interference from pilot contamination (see first term in the denominator of \eqref{eq:SINR_rp_gen}). This approach is simple and provides good results when a pilot reuse factor is used \cite{T_Marzetta_Ma_MIMO_book}. Thus, it provides a suitable comparison reference when evaluating the performance of SP. The selection of $\tau_p$ is of paramount importance in order to assess the performance of RP. Therefore, in Section~\ref{sec:num_result} we provide numerical results when $\tau_p$ is optimized to maximize the data rates. This optimization is done through an exhaustive search over the integer values of $\tau_p\in[K,\tau_c]$.  }
\subsection{Superimposed pilots}
In the case of SP, the received signal for an arbitrary data symbol $j$ in the coherence block, at $\BS_{0}$, is given by the $j^{th}$ column of $\Z_{0k}^\textsc{sp}$ (see \eqref{eq:rec_sig_pilot_SIP}). By combining the received signal $[\Z_{0k}^\textsc{sp}]_j$ with $\v_{00k}$, an estimate of the data symbol $j$ transmitted by $\UE_{0k}$ is obtained as ${[\hat{\s}_{0k}]_j  = \v_{00k}^H\left[\Z_{0k}^\textsc{sp}\right]_j}$. To compute an ergodic achievable rate, we first isolate the term that contains the desired information. To this end, we rewrite the detector as
\begin{equation}
\v_{00k} =  \upsilon_{00k}\bar{\gamma}_{0k}^\textsc{sp}  \h_{00k} + \bar{\v}_{00k} = \frac{1}{\sqrt{ M\beta_{00k}}} \h_{00k} + \bar{\v}_{00k}
\end{equation}
where
\begin{align}
\bar{\v}_{00k} &= \frac{\upsilon_{00k}\bar{\gamma}_{0k}^\textsc{sp}}{\sqrt{q_{0k} \tau_c}} \left(  \sum_{l'\in\Psi_{D}}\sum_{i=1}^{K} \rchi_{l'i}^\textsc{sp} \sqrt{q_{l'i}\tau_c} \h_{0l'i}  + \sum_{l'\in \Phi_{D}} \sum_{i=1}^{K} \sqrt{\frac{p_{l'i}}{\tau_c}} \h_{0l'i}\s_{l'i}^T\bvphi_{0k}^* +   \sum_{j'=1}^{\tau_c} \n_{0j'}\frac{\left[\bvphi_{0k}\right]_{j'}^*}{\sqrt{\tau_c}}\right).
\end{align}
Next, we add and subtract $\sqrt{\frac{p_{0k}}{M \beta_{00k}}}\E\left\{\left\|\h_{00k}\right\|^2 \right\} [\s_{0k}]_j$ from the data estimate $[\hat{\s}_{0k}]_j $ to obtain a desired signal with deterministic effective channel gain. This leads to  
\begin{equation}
\begin{aligned}
\label{eq:est_data_sip}
\left[ \hat{\s}_{0k} \right]_j 
&= \sqrt{\frac{p_{0k}}{M \beta_{00k}}}\E\left\{\left\|\h_{00k}\right\|^2 \right\} [\s_{0k}]_j + \sqrt{\frac{p_{0k}}{M \beta_{00k}}}\left( \left\|\h_{00k}\right\|^2 - \E\left\{\left\|\h_{00k}\right\|^2 \right\}\right)[\s_{0k}]_j\\
&\mkern 10mu + \underbrace{\sqrt{p_{0k}}\bar{\v}_{00k}^H\h_{00k}[\s_{0k}]_j
+ \sum_{l'\in \Phi_{D}} \sum_{i=1}^{K} \left( \sqrt{q_{l'i}}[\bvphi_{l'i}]_j +  \xi_{l'i}\sqrt{p_{l'i}} [\s_{l'i}]_j\right)\v_{00k}^H\h_{0l'i}   + \v_{00k}^H\    \n_{0j}}_{=n_\textit{eff}}.
\end{aligned}
\end{equation}
The term $n_\textit{eff}$ is defined in \eqref{eq:est_data_sip} for analytical tractability and accounts for the interference caused by pilot and data symbols received from all UEs (including self-interference from $\UE_{0k}$) plus noise. For ease of notation, we define $\xi_{l'i} = 0$ for $\{l',i\}  = \{0, k\}$ and $\xi_{l'i} = 1$ otherwise.

Notice that the first term in \eqref{eq:est_data_sip} is uncorrelated with the remaining ones in \eqref{eq:est_data_sip} since the data symbols have zero mean, are independent and circularly symmetric complex Gaussian. Thus, we have an equivalent SISO system with deterministic effective channel and non-Gaussian effective noise for which we can obtain an achievable rate based on the analysis in \cite[Ch. 2]{T_Marzetta_Ma_MIMO_book}. This result is summarized in the following theorem.
 {\color{black}\begin{theorem}
 	\label{th:ach_rate_ssf_SIP}
	An ergodic achievable rate for $\UE_{0k}$ with SP and MRC detection is
	\begin{equation}
	\AchR_{0k}^\textsc{sp} = \Bw \log_2\left(1 + \SINR_{0k}^\textsc{sp}\right)
	\label{eq:rate_sp_gen}
	\end{equation}
	where $\SINR_{0k}^\textsc{sp}$ is the effective SINR of $\UE_{0k}$ given by
 	\begin{align*} \allowdisplaybreaks
 	&\SINR_{0k}^\textsc{sp}  
 	= \frac{\frac{p_{0k}}{M \beta_{00k}}\left|\E\left\{\left\|\h_{00k}\right\|^2 \right\}\right|^2}{\frac{p_{0k}}{M \beta_{00k}} \left(\E\left\{  \left\|\h_{00k}\right\|^4\right\} -\left|\E\left\{ \left\|\h_{00k}\right\|^2 \right\}\right|^2 \right)  + \E\left\{ \left|n_\textit{eff} - \E\{n_\textit{eff}\}\right|^2\right\}}
 	\stepcounter{equation}\tag{\theequation}\label{eq:SINR_SIP_0k}
	\\
	&= M p_{0k}\beta_{00k}\! \Bigg/\! \Bigg(\!
	\underbrace{\!\frac{M}{\tau_c}\!\!\sum_{l'\in \Psi_{D}}\!\sum_{i =1}^{K}\!\frac{\left(\!p_{l'i}\!+\!\left(\!1\! -\! \frac{1}{\tau_c}\!\right)\!q_{l'i}\!\right)\!q_{l'i}}{q_{0k}}\frac{\beta_{0l'i}^2}{\beta_{00k}} \!+\! \frac{M}{\tau_c}\!\!\sum_{l'\in \Phi_{D}}\!\sum_{i =1}^{K}\!\frac{\left(\!p_{l'i}\! +\!q_{l'i}\!\right)\! p_{l'i} }{q_{0k}}\frac{\beta_{0l'i}^2}{\beta_{00k}}}_{\text{Coherent interference}}\\
	&\mkern 0mu + \! \underbrace{\!\frac{2}{\tau_c} p_{0k}\beta_{00k}\!+\! \frac{2}{\tau_c^2} \!\sum_{l'\in \Psi_{D}}\!\sum_{i =1}^{K}\!\frac{q_{l'i}p_{l'i}}{q_{0k}}\!\frac{\beta_{0l'i}^2}{\beta_{00k}}
		\!+\!   \frac{1}{\tau_c^2 }\!\sum_{l'\in \Phi_{D}}\!\sum_{i=1}^{K} \!\frac{p_{l'i}^2}{q_{0k}}\!\frac{\beta_{0l'i}^2}{\beta_{00k}} \!+\!\frac{1}{\gamma_{0k}^\textsc{sp}}\left( \sum_{l'\in\Phi_{D}}\!\sum_{i=1}^{K}\!\left(q_{l'i} \!+\! p_{l'i}\right)\! \beta_{0l'i} \!+\!\sigma^2\right)\!}_{\text{Non-coherent interference and noise}}\Bigg)  	
 	\stepcounter{equation}\tag{\theequation}\label{eq:SINR_sip_gen}
 	\end{align*}
where
 	\begin{align}
		\gamma_{0k}^\textsc{sp} &= \E\left\{\frac{1}{\bar{\gamma}_{0k}^\textsc{sp}} \right\}^{-1}= \frac{q_{0k}\tau_c \beta_{00k}}{q_{0k} \tau_c\beta_{00k} +  \sum\limits_{l'\in\Psi_{D}} \sum\limits_{i=1}^{K} q_{l'i} \beta_{0l'i}  +\sum\limits_{l'\in\Phi_{D}} \sum\limits_{i =1}^K \mkern 0mu p_{l'i}\beta_{0l'i} + \sigma^2}.
		\label{eq:gamma_sip_gen}
 	\end{align}
 	The term $n_\textit{eff}$ contains the last terms of the effective noise defined in \eqref{eq:est_data_sip}. The expectations in \eqref{eq:SINR_SIP_0k} are taken with respect to the SSF and the random pilot allocation.
 \end{theorem}}
\begin{IEEEproof}
	It follows from taking the estimate of $\left[ \hat{\s}_{0k} \right]_j$ in \eqref{eq:est_data_sip} and establishing an equivalent SISO system with a deterministic channel and uncorrelated non-Gaussian effective noise. Then, by applying standard lower bounds on the capacity between the transmitter and receiver of the equivalent SISO system, the ergodic achievable rate with effective SINR shown in \eqref{eq:SINR_SIP_0k} is derived \cite[Ch. 2]{T_Marzetta_Ma_MIMO_book}. The proof for obtaining the closed-form expression in \eqref{eq:SINR_sip_gen} can be found in Appendix~\ref{app:th_1}.
\end{IEEEproof}
With SP, there is no pre-log factor in \eqref{eq:rate_sp_gen} since the whole coherence block is used for data transmission. The coherent gain (see the numerator of \eqref{eq:SINR_sip_gen}) scales with $M$ and depends on the factor $\gamma_{0k}^\textsc{sp}$ (see \eqref{eq:gamma_sip_gen}), which reflects the channel estimation quality.  We define the coherent interference as the interference that adds constructively in the detection process due to the correlation between the detection vector and the received signal. As a result, its variance scales with $M$. With non-coherent interference, we refer to all the sources of interference that are combined non-constructively whose variance, in turn, does not scale with $M$. There is coherent interference from pilot contamination and also from pilot and data symbols (see the first two terms in the denominator of \eqref{eq:SINR_sip_gen}) due to the correlation between channel estimates and data symbols. Similarly, there is non-coherent interference from pilot symbols, data symbols and cross-correlation of the two (see the third and fourth terms in the denominator of \eqref{eq:SINR_sip_gen}). {\color{black}In the prior works \cite[Eq.~(12)]{SIP_part1_KU_SA}  and  \cite[Eq.~(41)]{VT_SP_approx_2016}, approximate SINR expressions are provided with SP and MRC based on asymptotic favorable propagation and channel hardening (i.e., $\lim\limits_{ M \to \infty} \frac{\h_{0li}^H\h_{0l'i'}}{M} = 0$ if $\{l,i\}\neq \{l',i'\}$ and $\lim\limits_{ M \to \infty} \frac{\|\h_{0li}\|^2}{M} = \beta_{0li}$). In contrast, the result in Theorem~\ref{th:ach_rate_ssf_SIP} does not rely on any asymptotic approximation. This enables us to accurately analyze the system performance for any finite $M$. By comparing \cite[Eq.~(12)]{SIP_part1_KU_SA} and \cite[Eq.~(41)]{VT_SP_approx_2016} with \eqref{eq:SINR_sip_gen}, it is seen that \eqref{eq:SINR_sip_gen} contains extra interfering terms, which might greatly affect the system performance.   }

{\color{black}Notice that since the pilot symbols are known to the BSs, they can be subtracted from $n_\textit{eff}$ to reduce the interference and obtain a better estimate of data symbols \cite{SIP_part1_KU_SA}. To obtain clear insights into the effect of the interference from pilot symbols, suppose the received pilot symbols can be perfectly removed from $n_\textit{eff}$. Let
		\setcounter{equation}{26}\begin{align}
		\bar{n}_\textit{eff} = \sqrt{p_{0k}}\bar{\v}_{00k}^H\h_{00k}[\s_{0k}]_j
		+ \sum_{l'\in \Phi_{D}} \sum_{i=1}^{K}  \xi_{l'i}\sqrt{p_{l'i}} [\s_{l'i}]_j\v_{00k}^H\h_{0l'i}   + \v_{00k}^H\n_{0j}
		\end{align}
		be the resulting term without the effect of pilot interference. Then, by replacing $n_\textit{eff}$ with $\bar{n}_\textit{eff}$ in \eqref{eq:SINR_SIP_0k} allows to computed an upper bound on the effective SINR with SP. This is summarized in the following corollary.
		\begin{corollary}\label{cor:SP_SINR_UB}
			By removing the received pilot symbols perfectly from the data estimates, the effective SINR with SP is upper bounded as $\SINR_{0k}^\textsc{sp}\leq\SINR_{0k}^\textsc{sp-ub}$ where
			\begin{align*} \allowdisplaybreaks
			&\SINR_{0k}^\textsc{sp-ub}  		
			=  M p_{0k}\beta_{00k}\! \Bigg/\! \Bigg(\!
			\underbrace{\!\frac{M}{\tau_c}\!\!\sum_{l'\in \Psi_{D}}\!\sum_{i =1}^{K}\!\frac{p_{l'i}q_{l'i}}{q_{0k}}\frac{\beta_{0l'i}^2}{\beta_{00k}} \!+\! \frac{M}{\tau_c}\!\!\sum_{l'\in \Phi_{D}}\!\sum_{i =1}^{K}\!\frac{p_{l'i}^2 }{q_{0k}}\frac{\beta_{0l'i}^2}{\beta_{00k}}}_{\text{Coherent interference}}\\
			&\mkern 200mu +  \underbrace{\frac{1}{\tau_c^2 }\sum_{l'\in \Phi_{D}}\sum_{i=1}^{K} \frac{p_{l'i}^2}{q_{0k}}\frac{\beta_{0l'i}^2}{\beta_{00k}} + \frac{1}{\gamma_{0k}^\textsc{sp}}\left(\sum_{l'\in\Phi_{D}}\sum_{i=1}^{K} p_{l'i}\beta_{0l'i} +\sigma^2\right)}_{\text{Non-coherent interference and noise}}  \Bigg) 	 	\label{eq:SINR_sip_gen_perf_pilot_sub}
			\stepcounter{equation}\tag{\theequation}
			\end{align*}		
		\end{corollary}
		\begin{IEEEproof}
			It follows from replacing $n_\textit{eff}$ with $\bar{n}_\textit{eff}$ in \eqref{eq:SINR_SIP_0k} and deriving the closed-form expression with the same approach as in Appendix~\ref{app:th_1}.
		\end{IEEEproof}By subtracting the received pilot symbols perfectly from $n_{\textit{eff}}$, both the coherent and non-coherent interference are reduced and some of the cross terms in the non-coherent interference vanish. This can increase the data rates provided that the proportion of power used for pilot symbols is not negligible. However, in practice the pilot symbols cannot be perfectly removed from data estimates because channels are not perfectly known (see Section~\ref{subsec:ch_est_SP}). Alternatively, we can remove the estimates of the received pilot symbols (i.e., $\sum_{l'\in\Phi_{D}}\sum_{i=1}^{K}\sqrt{q_{l'i}}[\bvphi_{l'i}]_j \v_{00k}^H\hat{\h}_{0l'i}$) from $n_{\textit{eff}}$. This approach would introduce a large number of cross terms into variance of $n_{\textit{eff}}$ since the channel estimates are correlated with the received data symbols of all UEs (see Remark~\ref{rem:ch est_SP}), and a closed-form expression of the effective SINR would not provide clear insights into the performance. The effect of removing the estimates of the received pilot symbols is evaluated numerically in Section~\ref{sec:num_result}.}

{\color{black}Notice that iterative decoding algorithms can be used to improve channel and data estimates. This is achieved at the price of an increased computational complexity with SP since the number of operations in each iteration grows linearly with $M$ and $\tau_c$ \cite{SIP_part1_KU_SA}. Moreover, similar approaches can also be used with RP where the data estimates can be used to improve the channel estimates and vice versa. As the first capacity analysis with SP, we focus on MRC detection and use the results with perfect pilot subtraction (shown in Corollary~\ref{cor:SP_SINR_UB}) to evaluate the possible gains of more complex signal processing schemes. The use of iterative decoding algorithms is thus left for future work.
	}

  \begin{sidewaystable}[ph!]%C{12cm}!t
{\color{black}
  	\small{
  		\renewcommand{\arraystretch}{1.3}
  		\caption{Achievable rate comparison of RP and SP}
%  		\hspace*{-0mm}
  		\label{tab:ach_rate_comp}
  		\centering
  		\begin{tabular}{m{0.21\textwidth}|C{0.128\textwidth}|C{0.12\textwidth}|C{0.21\textwidth}|C{0.23\textwidth}}
  			\bfseries Term & \bfseries RP Lemma~\ref{lem:ach_rate_ssf_RP} & \bfseries RP Theorem~\ref{th:LB_ach_rate_wrt_d} & \bfseries SP Theorem~\ref{th:ach_rate_ssf_SIP} and Corollary~\ref{cor:SP_SINR_UB} & \bfseries SP Theorem~\ref{th:LB_ach_rate_wrt_d} \\%[-2pt]
  			\hline
  			Coherent gain: Numerator of \eqref{eq:SINR_rp_gen}, \eqref{eq:SINR_sip_gen}, \eqref{eq:SINR_sip_gen_perf_pilot_sub}, \eqref{eq:SINR_LB_rate_lsf_rp}, \eqref{eq:SINR_LB_rate_lsf_sp} and \eqref{eq:SINR_LB_rate_lsf_sp_UB}.	
  			& 	
  			
  			$M p_{0k} \beta_{00k} $
  			
  			&
  			
  			$M $
  			
  			&
  			
  			$M p_{0k}\beta_{00k}$
  			
  			&
			$M $  			  		  			
  			
  			\\ \hline
  			Pilot contamination: coherent interference from UEs using the same pilot as $\UE_{0k}$				
  			&
  			\vspace*{-7mm}
  			$\frac{M}{\tau_p}\!\! \sum\limits_{l'\in\Psi_{D}}\!\sum\limits_{i=1}^{K}\!\frac{p_{l'i}q_{l'i}\beta_{0l'i}^2}{q_{0k}\beta_{00k}}$
  			&

  			$\frac{M K}{\tau_p\left(\alpha - 1\right)} $
  			
  			&  			
  			\vspace*{-8mm}
  			\begin{tabular}{L{0.21\textwidth}}
  			$\mkern -20mu$	No pilot subtraction:	\\%[-4mm]	
  			$\mkern -20mu$	$\frac{M }{\tau_c}\!\sum\limits_{l'\in \Psi_{D}}\!\sum\limits_{i =1}^{K}\!\frac{\left(p_{l'i} + \left(\!1 - \frac{1}{\tau_c}\!\right)q_{l'i}\right) q_{l'i}}{q_{0k}}\frac{\beta_{0l'i}^2}{\beta_{00k}}$	\\[-2mm]  			
  			$\mkern -20mu$	Perfect pilot subtraction:\\[-1mm]				
  			$\mkern -20mu$	$\frac{M }{\tau_c}\!\sum\limits_{l'\in \Psi_{D}}\!\sum\limits_{i =1}^{K}\!\frac{p_{l'i} q_{l'i}}{q_{0k}}\frac{\beta_{0l'i}^2}{\beta_{00k}}$\\[1mm] 
%  			$a_{l'i} = p_{l'i}$			
  			\end{tabular}
  			&		  				   				  			
  			\vspace*{-10mm}
  			\begin{tabular}{m{0.23\textwidth}}
			$\mkern -20mu$	No pilot subtraction:\\[-1mm]
			$\mkern -20mu$	$\frac{M K\left(1 - \frac{\Delta}{\tau_c}\right)}{\tau_c(\alpha - 1)}$ \\[-2mm]	
			$\mkern -20mu$	Perfect pilot subtraction:\\[-2mm]
			$\mkern -20mu$	$\frac{M K(1 - \Delta)}{\tau_c(\alpha - 1)}$
  			\end{tabular}
  			\\ \hline	
  			Additional coherent interference
  			&
  			
  			&		
  			&
  			\vspace*{-8mm}
%  			\hspace*{-6mm}  			
%  			\hskip-3mm
  			\begin{tabular}{L{0.21\textwidth}}
  			$\mkern -20mu$	No pilot subtraction:	\\[-1mm]	
  			$\mkern -20mu$	$\frac{M }{\tau_c}\!\sum\limits_{l'\in \Psi_{D}}\!\sum\limits_{i =1}^{K}\!\frac{ (p_{l'i} + q_{l'i})p_{l'i}}{q_{0k}}\frac{\beta_{0l'i}^2}{\beta_{00k}}$	\\[-1mm]  			
  			$\mkern -20mu$	Perfect pilot subtraction:\\[-1mm]				
  			$\mkern -20mu$	$\frac{M }{\tau_c}\!\sum\limits_{l'\in \Psi_{D}}\!\sum\limits_{i =1}^{K}\!\frac{ p_{l'i}^2}{q_{0k}}\frac{\beta_{0l'i}^2}{\beta_{00k}}$
%  			$a_{l'i} = p_{l'i}$			
  			\end{tabular}
 			&
  			\vspace*{-10mm}
%  			\hspace*{-6mm}  			
  			%  			\hskip-3mm
  			\begin{tabular}{m{0.23\textwidth}}
  			$\mkern -20mu$	No pilot subtraction:\\[-2mm]			
  			$\mkern -20mu$	$  \frac{M K (1 - \Delta)\alpha}{\tau_c\Delta (\alpha - 1)}$\\[-2mm]	
  			$\mkern -20mu$	Perfect pilot subtraction:\\[-2mm]				
  			$\mkern -20mu$	$ \frac{M K(1 - \Delta)^2\alpha}{\tau_c\Delta(\alpha - 1)} $			
  			\end{tabular}
  			
  			\\	\hline
  			
  			Non-coherent interference
  			&
  			$\left(\sum\limits_{l'\in\Phi_{D}}\! \sum\limits_{i=1}^{K}\!p_{l'i}\beta_{0l'i}\right)\frac{1}{\gamma_{0k}^\textsc{rp}}$
  			&
  			$\frac{K^2 }{\tau_p \left(\alpha - 1\right)}$ $+ 
  			\left( \frac{\alpha K }{\alpha - 2}\right)$ 
  			$\cdot \left(1 + \frac{K}{\tau_p} \frac{2}{\alpha - 2} + \frac{\sigma^2 }{\rho \tau_p} \right)$  
  			
  			&
  			\vspace*{-5mm}
%  			\hspace*{-6mm}  			
%  			\hskip-3mm
			\begin{tabular}{L{0.21\textwidth}}
				$\mkern -15mu$	$\underbrace{\!\!\frac{2}{\tau_c}a_{0k}\beta_{00k}\!+\! \frac{2}{\tau_c^2} \!\sum\limits_{l'\in \Psi_{D}}\!\sum\limits_{i=1}^{K}\!\frac{q_{l'i}a_{l'i}}{q_{0k}}\!\frac{\beta_{0l'i}^2}{\beta_{00k}}\!\!}_{\substack{\\[-2mm]\text{Cross products pilots and data}}}$\\%[-1mm]			
				$\mkern -15mu$	$+\frac{1}{\tau_c^2 }\sum\limits_{l'\in \Phi_{D}}\!\sum\limits_{i=1}^{K} \!\frac{p_{l'i}^2}{q_{0k}}\!\frac{\beta_{0l'i}^2}{\beta_{00k}}$	\\%[-1mm]			
				$\mkern -15mu$	$+\left(\sum\limits_{l'\in \Phi_{D}}\sum\limits_{i=1}^{K}(b_{l'i} + p_{l'i}) \beta_{0l'i}\right)\frac{1}{\gamma_{0k}^\textsc{sp}}$\\[-2mm]
				$\mkern -15mu$	No pilot subtraction:	\\[-4mm]	
				$\mkern -15mu$	$a_{l'i} = p_{l'i}$, $ b_{l'i} = q_{l'i}$ \\[-3mm]	
				$\mkern -15mu$	Perfect pilot subtraction:\\[-4mm]				
				$\mkern -15mu$	$a_{l'i} = b_{l'i} = 0$			
			\end{tabular}
  			&  			
  			\vspace*{-7mm}
  			\begin{tabular}{L{0.23\textwidth}}
  				$\mkern -15mu$	$\underbrace{\!\!\frac{2a}{\tau_c}\!\left(\!1\! +\! \frac{K}{\tau_c\left(\alpha \!-\! 1\right)}\!\right)\!\!}_{\text{Cross products pilots and data}}\mkern -18mu+ \frac{K(1\! -\! \Delta)^2 \alpha}{\tau_c^2\Delta \left(\alpha \!- \!1\right)} \!+\! \frac{K^2 b}{\tau_c \Delta \left(\alpha \!-\! 1\right)}$
  				\\%[-1mm]				
  				$\mkern -15mu$	$+\frac{K \alpha b}{\alpha \!-\! 2} \! \left(\!1\! +\! \frac{K}{\tau_c \Delta } \!\left(\!\frac{2  }{\left(\alpha \!-\! 2\right)} \!+\!(1 \!-\! \Delta)\!\right) \!+\! \frac{\sigma^2}{ \Delta \rho \tau_c} \!  \right)$\\
  				$\mkern -15mu$	No pilot subtraction:	\\[-4mm]	
  				$\mkern -15mu$	$a = 1 - \Delta$, $ b  = 1$ \\[-3mm]	
  				$\mkern -15mu$	Perfect pilot subtraction:\\[-4mm]				
  				$\mkern -15mu$	$a  = 0$, $b  = 1 - \Delta$			
  			\end{tabular}
  			\\				
  			\hline
  			%%\Xhline{4\arrayrulewidth}
  		\end{tabular}
  		%	\vspace*{-3mm}
  	}}
  \end{sidewaystable}
\section{Analysis of achievable rates}
\label{sec:ach_rate_analysis}
To compare the rate expressions in Lemma~\ref{lem:ach_rate_ssf_RP}, Theorem~\ref{th:ach_rate_ssf_SIP} {\color{black}and Corollary~\ref{cor:SP_SINR_UB}}, we characterize the terms in the effective SINR expressions \eqref{eq:SINR_rp_gen} with RP and \eqref{eq:SINR_sip_gen}{\color{black}, \eqref{eq:SINR_sip_gen_perf_pilot_sub}}  with SP, and analyze their influence on the network performance. From Table \ref{tab:ach_rate_comp}, we can see that by using the full coherence block for pilots in SP: \emph{i}) the estimates improve when $\tau_c$ increases; \emph{ii}) there is no penalty in the pre-log factor on the achievable rate; and \emph{iii}) the pilot contamination is reduced by a factor of {\color{black}$1/\tau_c$}. However, due to the high correlation between the received signal $[\Z_{0k}^\textsc{sp}]_j$ and the channel estimate $\hat{\h}_{00k}$, there are other interfering terms that are combined coherently or non-coherently. {\color{black}By subtracting perfectly the received pilot symbols, the coherent and non-coherent interference is reduced and several interference terms from cross products between pilot and data symbols vanish.} The relative strengths of the {\color{black}interference} terms depend on the network deployment setup. 

To gain further insights, we consider the asymptotic limit when $M \rightarrow \infty$. This shows the influence of the interference that combines coherently in the detection process. The asymptotic limits are summarized in the following corollary.
\begin{corollary}
	The achievable rates of $\UE_{0k}$ with RP and SP when $M\to \infty$ are given by
	{\color{black}
	\begin{align*}
	\stepcounter{equation}\tag{\theequation}\label{eq:SINR_rp_M_inf}
	\AchR_{0k}^\textsc{a-rp} &= \left(1 - \frac{\tau_p}{\tau_c}\right)\Bw\log_2\left(1+  \frac{ p_{0k} \beta_{00k} }{ \frac{1}{\tau_p} \sum\limits_{l'\in\Psi_{D}}\sum\limits_{i=1}^{K} \frac{p_{l'i} q_{l'i}}{q_{0k}}\frac{\beta_{0l'i}^2}{\beta_{00k}}}\right)\\
	\AchR_{0k}^\textsc{a-sp}&=\Bw\log_2\left(1 + \frac{p_{0k}\beta_{00k}}{ 
		\!\frac{1}{\tau_c}\!\sum\limits_{l'\in \Psi_{D}}\!\sum\limits_{i =1}^{K}\!\frac{\left(\!p_{l'i}+\left(1 - \frac{1}{\tau_c}\right)q_{l'i}\right)q_{l'i}}{q_{0k}}\frac{\beta_{0l'i}^2}{\beta_{00k}} \!+\! \frac{1}{\tau_c}\!\sum\limits_{l'\in \Phi_{D}}\!\sum\limits_{i =1}^{K}\!\frac{\left(p_{l'i} +q_{l'i}\right)p_{l'i} }{q_{0k}}\frac{\beta_{0l'i}^2}{\beta_{00k}}}	
		\right)	\stepcounter{equation}\tag{\theequation}\label{eq:SINR_sip_M_inf}\\
		&\leq\Bw\log_2\left(1 + \frac{p_{0k}\beta_{00k}}{ 
			\!\frac{1}{\tau_c}\!\sum\limits_{l'\in \Psi_{D}}\!\sum\limits_{i =1}^{K}\!\frac{p_{l'i}q_{l'i}}{q_{0k}}\frac{\beta_{0l'i}^2}{\beta_{00k}} \!+\! \frac{1}{\tau_c}\!\sum\limits_{l'\in \Phi_{D}}\!\sum\limits_{i =1}^{K}\!\frac{p_{l'i}^2 }{q_{0k}}\frac{\beta_{0l'i}^2}{\beta_{00k}}}	
		\right).
	\stepcounter{equation}\tag{\theequation}\label{eq:SINR_sip_M_inf_sub_perf_pilot}
	\end{align*}}
	\label{cor:asymptotic_M_infty}
\end{corollary}
\begin{IEEEproof}
It follows from taking the limit in the expressions \eqref{eq:SINR_rp_gen}, \eqref{eq:SINR_sip_gen} {\color{black}and \eqref{eq:SINR_sip_gen_perf_pilot_sub}}. 
\end{IEEEproof}
The above asymptotic formulas can be used to compare RP and SP. We see that the scaling factor $1/\tau_p$ in the coherent pilot contamination with RP (see the denominator of the fraction inside the logarithm in \eqref{eq:SINR_rp_M_inf}) is larger than $1/\tau_c$ with SP (see the denominator of the fraction inside the logarithm in \eqref{eq:SINR_sip_M_inf}{\color{black}, \eqref{eq:SINR_sip_M_inf_sub_perf_pilot}}). However, with SP there is another term with coherent interference that affects the performance. If we compare the results in Corollary~\ref{cor:asymptotic_M_infty} with \mbox{\cite[Eq.~(13)]{SIP_part1_KU_SA}} {\color{black} and \cite[Eq.~(42)]{VT_SP_approx_2016}}, the following two differences are observed. First, the pilot contamination term with SP is neglected in \cite{SIP_part1_KU_SA, VT_SP_approx_2016}, which is a valid assumption only for scenarios wherein the total number of UEs in the entire network is lower than $\tau_c$. However, this is not the case of practical networks with many cells, and thus there will be pilot contamination also with SP. Secondly, in \cite{SIP_part1_KU_SA} the length of the pilot signals $\tau_p$ with RP is assumed not to change with the coherence block size. However, this parameter can indeed be optimized for a given size of the coherence block. As a result, with RP the effect of pilot contamination changes with the size of the coherence block as well and this could affect the scenarios in which SP outperforms RP, and vice versa.

To compare the asymptotic achievable rates given by \eqref{eq:SINR_rp_M_inf} and \eqref{eq:SINR_sip_M_inf}{\color{black}, \eqref{eq:SINR_sip_M_inf_sub_perf_pilot}}, we define $\zeta = \tau_p/\tau_c$ as the ratio between the pilot length with RP and the size of the coherence block. The value of $\zeta$ that maximizes the achievable rate with RP is given as follows. 
\begin{corollary}	
	With RP, the asymptotic rate when $M\to \infty$ is a concave function of $\zeta$ and its maximum is found at
	\begin{align}
	\zeta^\texttt{max} =  \frac{1}{\SIR_\textsc{rp}} \left( \frac{1 + \SIR_\textsc{rp}}{W\left(\left(1 + \SIR_\textsc{rp}\right) e \right)} - 1\right) \in (0,1)
	\label{eq:zeta_min}
	\end{align}
	for $\SIR_\textsc{rp} >0$ where
	\begin{align*}
	\SIR_\textsc{rp} =  \frac{ p_{0k} \beta_{00k} }{ \frac{1}{\tau_c} \sum\limits_{l'\in\Psi_{D}}\sum\limits_{i=1}^{K} \frac{p_{l'i} q_{l'i}}{q_{0k}}\frac{\beta_{0l'i}^2}{\beta_{00k}}}	
	\end{align*}
	 and $W(\cdot)$ denotes the Lambert \textit{W} function\footnote{The Lambert \textit{W} function is defined as $z = W(z)e^{W(z)}$ for any $z\in \C$. More details can be found in \cite{Lambert_W}.} and $e$ denotes the base of the natural logarithm. 	
	\label{cor:realtion_rate_RP_SP}
\end{corollary}
\begin{IEEEproof}
The corollary is proved in Appendix \ref{app:relation_rate_RP_SP}.
\end{IEEEproof}
{\color{black} Notice that: \textit{i)} $\AchR_{0k}^\textsc{a-rp}$ is a concave function of $\zeta\in[0,1]$ that starts ($\zeta = 0$) and ends ($\zeta =1$) at zero and thus it is not monotonic; \textit{ii)} $\AchR_{0k}^\textsc{a-rp}$ depends linearly and logarithmically on $\zeta$. To the best of our knowledge, it is not possible to find in closed-form the solution to the inequality $\AchR_{0k}^\textsc{a-rp} \leq \AchR_{0k}^\textsc{a-sp}$ in terms of $\zeta$. Let $\AchR_{0k}^\textsc{a-rp-max} = \max_{\zeta\in [0,1]}\{\AchR_{0k}^\textsc{a-rp} \}$, then if $\AchR_{0k}^\textsc{a-sp} > \AchR_{0k}^\textsc{a-rp-max}$ the asymptotic achievable rate with SP always outperforms RP. However, if $\AchR_{0k}^\textsc{a-sp} < \AchR_{0k}^\textsc{a-rp-max}$ there exists an interval around $\zeta^\texttt{max}$ for which the asymptotic achievable rate with RP is better than SP. Since the inequality condition $\AchR_{0k}^\textsc{a-sp} > \AchR_{0k}^\textsc{a-rp}$ depends on the power allocation and LSF coefficients, we need to consider a particular network deployment setup to offer a more precise comparison between RP and SP. This is what we do in the next sections.}
\section{Performance Comparison under Random Deployment}
\label{sec:sto_geometry}
As shown in Sections~\ref{sec:ach_rates} and \ref{sec:ach_rate_analysis}, it is necessary to have a particular network deployment setup to obtain further insights into the performance of SP and RP. To model the irregularity and large number of cells of practical networks, we use the stochastic geometry framework from \cite{Emil_MAMIMO_small_cells}, which has been shown to accurately model real network deployments \cite{M_di_renzo_Exp_stoc_geometry_conf}. Here, the BSs are distributed according to a spatially homogeneous PPP, that is $\Phi_{D}$ is a homogeneous PPPs with density $D$ [BS/km$^2$]. Without loss of generality, we refer to $\BS_0$ as a \textit{typical} BS and to $\UE_{0k}$ as a \textit{typical} UE. Particularly, they represent any BS and UE in the network by means of the translation invariance property of the homogeneous PPP. The $K$ UEs in each cell are assumed to be uniformly distributed within the Voronoi region around each BS. This means that the distance between $\UE_{l'i}$ and $\BS_{l'}$, denoted by $d_{l'l'i}$ [km],  is distributed as $d_{l'l'i}\sim \mathrm{Rayleigh}\left(\frac{1}{\sqrt{2\pi D}} \right)$. To model the LSF between $\UE_{l'i}$ and $\BS_l$, we define $\beta_{ll'i} = \omega^{-1} d_{ll'i}^{-\alpha}$ where $\alpha$ is the pathloss exponent and $\omega$ is the pathloss at a reference distance of 1 km. This parameter also accounts for propagation losses independent of the distances (e.g., wall penetration).

In cellular networks, the transmission power of UEs needs to be controlled in order to avoid that signals from UEs close to the BS overwhelm signals from UEs further away. This is particularly important in Massive MIMO where low-resolution analog-to-digital converters are expected to be used \cite{Durisi_MAMIMO_low_res_ADC, C_Mollen_1bit_ADC}.  Thus, we assume statistical channel inversion power control where the transmission power of data symbols is computed as ${p_{l'i} =\rho_d /\beta_{l'l'i}= \rho_d \omega d_{l'l'i}^{\alpha}}$ and of pilot symbols as ${q_{l'i} = \rho_p/\beta_{l'l'i} = \rho_p \omega d_{l'l'i}^{\alpha}}$. The design parameters $\rho_d$ and $\rho_p$ are used to control the average transmit power for data and pilot symbols, respectively. Moreover, we define $\rho$ as the total average transmission power per symbol\footnote{{\color{black}The average SNR per UEs is then given by $\rm{SNR} = \rho/\sigma^2$.}} such that $\rho_d=\rho_p =\rho$ with RP and $\rho_d +\rho_p = \rho$ with SP. {\color{black}We define the proportion between pilot and data power with SP as $\Delta\in[0,1]$ such that $\rho_p = \Delta \rho$ and $\rho_d = (1 - \Delta) \rho$.}
 
{\color{black}By introducing the aforementioned definitions of transmission powers and LSF coefficients,} the achievable rates with RP and SP can be computed in terms of the distances between UEs and BSs. To get insights into the influence of design parameters such as the number of BS antennas $M$, the number of UEs per BS $K$, length of pilot sequences $\tau_p$ with RP and system parameters such as the size of the coherence block $\tau_c$,  we evaluate the performance for different realizations of the UE positions. In particular, we calculate an expected value of $\bar{\AchR}_{0k}^\textsc{rp}$ and $\bar{\AchR}_{0k}^\textsc{sp}$ with respect to the distances ${d_{ll'i}\;\forall l,l'\in \Phi_D}$ and ${i\in \{1,\ldots,K\}}$. Following the same approach as in \cite{Emil_MAMIMO_small_cells}, a closed-form lower bound on the achievable rates can be computed as shown in the following theorem.
{\color{black}
	\begin{theorem}
			\label{th:LB_ach_rate_wrt_d}	
	A lower bound on the average ergodic achievable rate of the \textit{typical} $\UE_{0k}$ with respect to the UE positions when considering statistical channel inversion power control is with RP given by
	\begin{align}
	\underline{\AchR}^\textsc{rp} &= \Bw \left(1 - \frac{\tau_p}{\tau_c}\right)\log_2\left(1 + \underline{\SINR}^\textsc{rp}\right)\\
	\underline{\SINR}^\textsc{rp}
	&= \frac{M  }{\frac{M K}{\tau_p\left(\alpha - 1\right)} + \frac{K^2 }{\tau_p \left(\alpha - 1\right)}+ 
		 \left( 1 + \frac{K}{\tau_p} \frac{2}{\alpha - 2} + \frac{\sigma^2 }{\rho \tau_p} \right)\left( \frac{\alpha K }{\alpha - 2} + \frac{\sigma^2}{\rho} \right)  	} 
	\label{eq:SINR_LB_rate_lsf_rp}
	\end{align}
	where $\rho_d = \rho_p = \rho$. With SP, it is given by 
	\begin{equation}	
	\underline{\AchR}^\textsc{sp} = \Bw \log_2\left(1 + \underline{\SINR}^\textsc{sp}\right)
	\end{equation}
	\begin{align*}\allowdisplaybreaks
	&\underline{\SINR}^\textsc{sp} =M(1 - \Delta) \Bigg/\Bigg(
	\underbrace{\frac{M K}{\tau_c\left(\alpha - 1\right)}\!\left(\!1 \!- \!\frac{\Delta}{\tau_c}\!\right) 
		\!+\! \frac{M K}{\tau_c}\frac{ \left(1 -\Delta\right) \alpha}{\Delta \left(\alpha - 1\right)}}_{\text{Coherent Interference}} 	
	\\
	&	
	\!+\! \underbrace{\!\frac{2(1\! - \!\Delta)}{\tau_c}\!\left(\!1\! +\! \frac{K}{\tau_c\left(\alpha \!-\! 1\right)}\!\right) 
	\!+\!   \frac{K(1\! -\! \Delta)^2 \alpha}{\tau_c^2\Delta \left(\alpha \!- \!1\right)} \!+\! \frac{K^2}{\tau_c \Delta \left(\alpha \!-\! 1\right)} \!+\!\left(\!1\! +\! \frac{K}{\tau_c \Delta } \!\left(\!\frac{2  }{\left(\alpha \!-\! 2\right)} \!+\!(1 \!-\! \Delta)\!\right) \!+\! \frac{\sigma^2}{ \Delta \rho \tau_c} \!  \right)\!\!\left(\!\frac{K \alpha }{\alpha \!-\! 2} \! +\!\frac{\sigma^2}{\rho} \! \right)\!\!\!}_{\text{Non-coherent Interference and noise}}\Bigg). 
		\stepcounter{equation}\tag{\theequation}
	\label{eq:SINR_LB_rate_lsf_sp}
	\end{align*}		
	By subtracting the pilot symbols perfectly from the data estimates, an upper bound on the effective SINR with SP is given by $\underline{\SINR}^\textsc{sp} \leq \underline{\SINR}^\textsc{sp-ub}$ where 
	\begin{align*}\allowdisplaybreaks
	\underline{\SINR}^\textsc{sp-ub} =&M(1 - \Delta) \Bigg /\Bigg(
		\underbrace{\frac{M K \left(1 \!-\! \Delta\right)}{\tau_c\left(\alpha\! -\! 1\right)} 
			\!+\! \frac{M K \left(1\! -\!\Delta\right)^2 \alpha}{\tau_c \Delta \left(\alpha\! - \!1\right)}}_{\text{Coherent Interference}}
		\\&
		\!+\! \underbrace{\frac{K(1 - \Delta)^2 \alpha}{\tau_p^2\Delta \left(\alpha - 1\right)} + \frac{K^2\left(1 - \Delta\right)}{\tau_p \Delta \left(\alpha - 1\right)}  \!+\!\left(\!1\! +\! \frac{K}{\tau_c \Delta } \!\left(\!\frac{2  }{\left(\alpha\! -\!2\right)} \!+\!(1\! -\! \Delta)\!\right) \!+\! \frac{\sigma^2}{ \Delta \rho \tau_c} \!  \right)\!\!\left(\!\frac{K \left(1\!-\!\Delta\right)\alpha }{\alpha \!-\! 2} \! +\!\frac{\sigma^2}{\rho} \! \right)\!\!}_{\text{Non-coherent Interference and noise}}\Bigg). 
	\stepcounter{equation}\tag{\theequation}
	\label{eq:SINR_LB_rate_lsf_sp_UB}
	\end{align*}
    For both SP and RP, $\rho$ is the average transmission power per symbol.	
\end{theorem}}
\begin{IEEEproof}
	It follows from applying Jensen's inequality to the achievable rate as {\color{black}\[\E\{\log_2( 1 + \SINR_{0k}) \}\geq \log_2(1 + 1/\E\{ \SINR_{0k}^{-1}\})\]} where {\color{black}$\SINR_{0k}$} represents the SINR of $\UE_{0k}$ with either RP or SP. We then compute the moments of {\color{black}$\SINR_{0k}^{-1}$}. Notice that the expectation with respect to the distances results in an SINR expression independent of the UE index ``$0,k$''. See Appendix \ref{app:proof_th_LB_ach_rate_wrt_d} for details on calculating {\color{black}$\E\{ \SINR_{0k}^{-1}\}$}.
\end{IEEEproof}
{\color{black}The lower bounds on} the achievable rates with both RP and SP are increasing with $M$ and decreasing with $K$, which means that in order to serve more UEs with the same rates we need to increase the number of BS antennas. With RP, we can see that $\underline{\SINR}^\textsc{rp}$ increases with $\tau_p$. However, the pre-log factor $\left(1 - \tau_p/\tau_c\right)$ decreases with $\tau_p$. {\color{black} This means that the rate $\underline{\AchR}^\textsc{rp}$ is a unimodal function of $\tau_p$ which can be maximized with bisection search algorithms}. This result is in line with Corollary~\ref{cor:realtion_rate_RP_SP}. If $\tau_c$ increases, then the pre-log factor $1 - \tau_p/\tau_c$ increases as well since $\tau_p \in [K,\tau_c]$. This means that, by optimizing $\underline{\AchR}^\textsc{rp}$ with respect to $\tau_p$, the maximum rate with RP increases with $\tau_c$. With SP, the achievable rate $\underline{\AchR}^\textsc{sp}$ also increases with $\tau_c$. 

{\color{black}Notice that the closed-form expressions found in Theorem \ref{th:LB_ach_rate_wrt_d} do not require heavy numerical simulations and can give powerful insights into the data rates of practical network deployments.} 

\subsection{Energy Efficiency Modeling}
\label{subsec:EE}
The closed-form achievable rates provided above to measure the SE of the network allow us to provide analytical expressions for the EE, measured in [bit/Joule], with RP and SP. We consider the effect of transmission and circuit power consumption following the model found in \cite{Emil_power_model}. We define the EE as the ratio between the average sum data rate per unit area and the average power consumption per unit area. This yields
\begin{align}
\EE =\frac{\E \left\{\sum_{k=1}^{K}{\color{black}\AchR_{0k}}\right\} D}{\Pbs D}= \frac{\E\left\{{\color{black}\AchR_{0k}}\right\} K}{\Ptx + \cC_0 + \cC_1 K + \cD_0 M + \Plp + \Pce + \cA \,\E\left\{{\color{black}\AchR_{0k}}\right\} K}
\label{eq:EE}
\end{align}	
where ${\color{black}\AchR_{0k}}$ is the achievable rate defined in {\color{black}\eqref{eq:rate_rp_gen}} with RP and {\color{black}\eqref{eq:rate_sp_gen}} with SP, and $\Pbs$ is the power consumption per BS, which accounts for the transmission power and circuit power consumption (see the denominator of the second equality of \eqref{eq:EE}). Recall that $D$ is the density of BSs per unit area in [BS/km$^2$]. Note that $\E\{{\color{black}\AchR_{0k}}\}$ can be replaced by its lower bound in Theorem~\ref{th:LB_ach_rate_wrt_d}. By using Lemma~\ref{lem:exp_d_alpha} in Appendix~\ref{app:stoc_geometry}, the average transmission power is computed as
\begin{equation}
\Ptx = \frac{\Bw}{\eta} K \E\left\{p_{0k}\right\}=\frac{\Bw}{\eta} K \rho \omega\frac{\Gamma\left( \alpha/2 + 1\right) }{\left(\pi D\right)^{\alpha/2}}
\end{equation}
where $\eta \in (0,1]$ denotes the efficiency of the power amplifier. The parameter $\cC_0$ accounts for fixed power consumption (e.g., site cooling), $\cC_1$ and $\cD_0$ are the power consumed per transceiver chain of the UE and BS, respectively. The power consumption that depends on the data rates (e.g., coding, decoding, backhaul, etc.) is enclosed by $\cA$. The power consumption for linear processing and channel estimation is denoted by $\Plp$ and $\Pce$, respectively. To calculate $\Plp$ and $\Pce$, we find a first-order approximation of the computational complexity (i.e., number of floating point operations per second (flops)), based on the number of complex multiplications in linear algebra operations, and then multiply it by the computational efficiency of current microprocessors denoted by $L$ [flops/W].  Then, we have
\begin{align}
\Plp + \Pce &= \begin{cases}
 \frac{\Bw}{L} M K & \text{ with RP}\\
2\frac{\Bw}{L}  M K & \text{ with SP.}
\end{cases}
\end{align}
The combined power consumption for linear processing and channel estimation is doubled by SP as compared to RP. This occurs because with SP we estimate the channel and detect the data for all symbols in the coherence block, whereas with RP we only estimate the channel in $\tau_p$ symbols and detect the data in $\tau_c-\tau_p$ symbols. Given that the computational efficiency of modern microprocessors is continuously increasing, a factor of two does not add a significant weight into the total power consumption with SP when compared to RP. Thus, the difference between RP and SP in terms of EE is going to be mainly determined by the rate performance. Comparisons are made in the next section.

\section{Numerical results}
\label{sec:num_result}
Monte Carlo (MC) simulations are used to compare RP and SP, and to validate the theoretical results of Sections \ref{sec:ach_rates}, \ref{sec:ach_rate_analysis} and \ref{sec:sto_geometry}. We simulate a homogeneous PPP with density $D = 100$ [BS/km$^2$] in an squared area of side length $L_{\textsc{sq}}$ [km] with an average of $N_{av} = D L_{\textsc{sq}}^2 = 50$ BSs. To avoid edge effects, we implement the wrap around technique where we replicate the original square 8 times and place the copies around itself. Table \ref{tab:sim_param} summarizes the simulation parameters which are based on \cite{Emil_MAMIMO_small_cells, Emil_power_model} and references therein.
\begin{table}[!t]%C{12cm}
	\renewcommand{\arraystretch}{1.3}
	\caption{Simulation Parameters}
	\label{tab:sim_param}
	\centering
	\begin{tabular}{c|c|c||c|c|c}
		\bfseries Parameter & \bfseries Symbol & \bfseries Value &\bfseries Parameter & \bfseries Symbol & \bfseries Value\\[-4pt]
		\hline
		%\Xhline{4\arrayrulewidth} %\hline
		 &  &  & &  & \\[-25pt]
		Pathloss exponent & $\alpha$ & $3.76$ & Circuit power per active UE & $\cC_1$ & $0.1$ W \\[-4pt]
		Fixed propagation loss (1 km) & $\omega$ & $130$ dB &Circuit power per BS antenna & $\cD_0$ & $0.1$ W \\[-4pt]
		System bandwidth & $\Bw$ & $20$ MHz & Power const. related to data rates & $\cA \Bw$ & $2.3\times 10^{-2}$ W\\[-4pt]
		Power amplifier efficiency & $\eta$ & 0.39 & Computational efficiency &   $L$ & $12.8$ [Gflops/W]\\[-4pt] %$4\times 10^{11}$
		Static power consumption & $\cC_0$ & 10 W & Noise power & $\sigma^2 \Bw$ &   $10^{-13}$ W  \\%[-4pt]
		\hline
		%%\Xhline{4\arrayrulewidth}
	\end{tabular}
\end{table}
We evaluate the performance of achievable rates and EE with MRC for the following methods: 
\begin{itemize}
	\item RP with pilot length equal to the number of users per BS, i.e., $\tau_p = K$;
	\item RP with optimal pilot length to maximize ${\color{black}\AchR_{0k}^\textsc{rp}}$;
	\item {\color{black}SP as in Theorem~\ref{th:ach_rate_ssf_SIP} (denoted as ``SP no sub.'', i.e., no pilot subtraction), Corollary~\ref{cor:SP_SINR_UB} (denoted as ``SP perf. sub.'',i.e., perfect pilot subtraction) and Theorem~\ref{th:LB_ach_rate_wrt_d}; %(denoted with the suffix ``L.B.'' that stands for lower bound);
	\item SP when we subtract the estimated received pilot symbols from the data estimate $[\hat{\s}_{0k}]_j$ in \eqref{eq:est_data_sip}, denoted as ``SP est. sub.'' which stands for estimated pilot subtraction;
	\item SP with the approximated results found in \cite{SIP_part1_KU_SA,VT_SP_approx_2016} denoted as ``Approx. \cite{SIP_part1_KU_SA}'' and ``Approx.  \cite{VT_SP_approx_2016}'' respectively.}
%	
%	 of UEs in cell $0$ to the  We only subtract estimates of received pilot symbols from UEs in cell $0$ since this provides the best result for SP transmission (see Remark~\ref{rem:sub_pilots}).\footnote{Note that the quality of the channel estimates of UEs in other cells is low, which is why we only subtract the estimates of received pilot symbols from UEs in cell $0$. These results were observed through numerical analysis and are omitted for brevity.}
\end{itemize}
 {\color{black} Note that in all figures the proportion $\Delta$ between pilot and data power with SP is optimized to maximize the data rates in each LSF realization.
% \begin{figure}[!t] %
% 	\centering %
%	 \begin{tikzpicture}
%	 \usetikzlibrary{plotmarks}
%	 \begin{axis}[
%	 width=8cm,
%	 height=6cm,
%	 xlabel={Number of antennas $M$},
%	 ylabel={Avg. UL rate per UE [Mbit/s]},
%	 ymin = 0,
%	 ymax = 100,
%	 line width = 1.2pt,
%	 grid=both,
%%	 legend pos= north west,
%	 legend style={at={(0.57,0.99)},font=\small,line width=0.5pt},%draw=none
%	 legend entries={SP no sub. pilot, ,  SP perf. sub. pilot,  , SP Approx. \cite{SIP_part1_KU_SA}, 'SP Approx. [27]'},
%	 ]% mark = none
%	 \addplot [only marks,mark=o, mark size=3pt,black] table[x index={0}, y index={1}] {Figures/SP_only_vs_M.txt};
%	 \addplot [solid,black,solid] table[x index={0}, y index={2}] {Figures/SP_only_vs_M.txt};
%	 \addplot [only marks,mark=triangle, mark size=4pt,green] table[x index={0}, y index={3},only marks] {Figures/SP_only_vs_M.txt};
%	 \addplot [solid,green]  table[x index={0}, y index={4}] {Figures/SP_only_vs_M.txt};
%	 \addplot [mark=x, mark size=4pt,magenta] table[x index={0}, y index={5}] {Figures/SP_only_vs_M.txt};
%	 \addplot [teal,mark=+, mark size=4pt] table[x index={0}, y index={6}] {Figures/SP_only_vs_M.txt};
%	 \end{axis}
%	 \end{tikzpicture}
% 	\caption{Ach. rate.} 
% 	\label{fig:_SP_comparison} 
% \end{figure} %

 \begin{figure*}[!t]
 	 \color{black}
 	\centering
 	{\captionsetup{width=0.4\textwidth}
 		\subfloat[Optimized parameters for $\rho = \sigma^2/4$ (SNR~${=-6}$ dB), $K = 10$ and $\tau_c =200$.]{
 			{\small 			
 				\begin{tabular}{C{45pt} C{30pt}C{30pt}C{30pt}}\\[-180pt] 	 
 					% 				\vspace*{-180pt}	
 					\hline			
 					\bfseries Param. &  $\!M\! =\! 100\!$ & $\!M\! =\! 300\!$ &$\!M \!= \!500\!$\\%[-2pt]
 					\hline
 					$\tau_p$ Opt. & 39 & 42 & 44 \\
 					\hline
 					$\Delta$ no sub. & 0.36 & 0.45& 0.5 \\
 					\hline
 					$\Delta$ perf sub. & 0.6 & 0.7 & 0.75 \\
 					\hline
	 				$\Delta$ est. sub. & 0.47 & 0.5 & 0.53 \\ 
	 				\hline
 					%%\Xhline{4\arrayrulewidth}
 				\end{tabular}}
 				\label{tab:opt_param}}
 			\hfil
 		\subfloat[Achievable rate per UE for $\rho = \sigma^2/4$ (SNR~${=-6}$ dB), $\tau_c = 200$ and $K =10$.]{
 		 	\begin{tikzpicture}
	\begin{axis}[
	width=0.42\textwidth,
	height=0.32\textwidth,
	y label style = {at={(-0.02,0.5)},anchor=north},
	xlabel={Number of antennas $M$},
	ylabel={Avg. rate per UE [Mbit/s]},
	ymin = 5,
	ymax = 102,
	xmin = 0,
	xmax = 500,
	ytick distance = 20,
%	xtick distance = 10,
	line width = 1.2pt,
	grid=both,
	legend columns = 3,	
	transpose legend,
	%	 legend pos= north west,
	legend style={at={(0.965,1)},font=\footnotesize,line width=1pt,draw=black,mark size=0.1pt},%draw=none font=\fontsize{4}{5}\selectfont %(0.64,0.99)%(0.515,0.99)
	]% mark = none		
	\addlegendentry{SP perf. sub.}
	\addlegendimage{solid,mark=triangle, mark size=4pt,green}

	\addlegendentry{Approx. \cite{SIP_part1_KU_SA}}
	\addlegendimage{densely dashed,mark=x, mark size=4pt, mark options={solid},magenta}

	\addlegendentry{SP est. sub.}
	\addlegendimage{only marks,mark=diamond, mark size=3pt,cyan}

	\addlegendentry{SP no sub.}
	\addlegendimage{solid,mark=o, mark size=3pt,blue}

	\addlegendentry{Approx. \cite{VT_SP_approx_2016}}	
	\addlegendimage{densely dashed,mark=+, mark size=4pt, mark options={solid, line width = 1.6pt},teal}

	\addlegendentry{}
	\addlegendimage{empty legend}

	\addplot [only marks,mark=o, mark size=3pt,blue] table[x index={0}, y index={1}] {Figures/SP_only_vs_M.txt};	
	\addplot [solid,blue,solid] table[x index={0}, y index={2}] {Figures/SP_only_vs_M.txt};
	\addplot [only marks,mark=triangle, mark size=4pt,green] table[x index={0}, y index={3},only marks] {Figures/SP_only_vs_M.txt};
	\addplot [solid,green]  table[x index={0}, y index={4}] {Figures/SP_only_vs_M.txt};
	\addplot [cyan,   only marks, mark=diamond, mark size=3pt] 	table[x index={0}, y index={5},only marks] {Figures/SP_only_vs_M.txt};		
	\addplot [magenta, densely dashed, mark=x, mark size=4pt, mark options={solid}] table[x index={0}, y index={6}] {Figures/SP_only_vs_M.txt};
	\addplot [teal, densely   dashed, mark=+, mark size=4pt, mark options={solid, line width = 1.6pt}] table[x index={0}, y index={7}] {Figures/SP_only_vs_M.txt};
	\end{axis}
% 		\draw (0.05,2.23) rectangle (3.66,4.1);
\end{tikzpicture}

% 'SP no sub. pilot'
% 'SP no sub. pilot'
% 'SP perf. sub. pilot'
% 'SP  perf. sub. pilot'
% 'SP est. sub. pilot'
% 'SP Approx. [25]'
% 'SP Approx. [27]'
 			\label{fig:R_only_SP_vs_M_SSF}}
 		 		\\[-12pt]
 		\subfloat[Achievable rate per UE for $\rho = \sigma^2/4$ (SNR~${=-6}$ dB), $M = 100$ and $K =10$.]{
 			\begin{tikzpicture}
	\begin{semilogxaxis}[
	width=0.42\textwidth,
	height=0.32\textwidth,
	y label style = {at={(-0.02,0.5)},anchor=north},%,font=\small
	xlabel={Size of coherence block $\tau_c$},
	ylabel={Avg. rate per UE [Mbit/s]},
	ymin = 0,
	ymax = 50,
	xmin = 10,
	xmax = 1000,
	ytick distance = 10,
	line width = 1.2pt,
	grid=both,
	%	 legend pos= north west,
	legend style={at={(0.46,0.99)},font=\footnotesize,line width=1pt,draw=black,mark size=0.1pt},%draw=none font=\fontsize{4}{5}\selectfont %(0.64,0.99)
	]% mark = none
	\addplot [only marks,mark=o, mark size=3pt,blue] table[x index={0}, y index={1}] {Figures/SP_only_vs_tau_c.txt};
	\addplot [solid,blue,solid] table[x index={0}, y index={2}] {Figures/SP_only_vs_tau_c.txt};
	\addplot [only marks,mark=triangle, mark size=4pt,green] table[x index={0}, y index={3},only marks] {Figures/SP_only_vs_tau_c.txt};
	\addplot [solid,green]  table[x index={0}, y index={4}] {Figures/SP_only_vs_tau_c.txt};
	\addplot [cyan,   only marks, mark=diamond, mark size=3pt] 	table[x index={0}, y index={5},only marks] {Figures/SP_only_vs_tau_c.txt};			
	\addplot [magenta,densely dashed, mark=x, mark size=4pt,mark options={solid}] table[x index={0}, y index={6}] {Figures/SP_only_vs_tau_c.txt};	
	\addplot [teal, densely   dashed, mark=+, mark size=4pt,mark options={solid, line width = 1.6pt}] table[x index={0}, y index={7}] {Figures/SP_only_vs_tau_c.txt};
	\end{semilogxaxis}
% 		\draw (0.05,2.23) rectangle (3.66,4.1);
\end{tikzpicture}

% 'SP no sub. pilot'
% 'SP no sub. pilot'
% 'SP perf. sub. pilot'
% 'SP  perf. sub. pilot'
% 'SP est. sub. pilot'
% 'SP Approx. [25]'
% 'SP Approx. [27]'
 			\hfil
 			\label{fig:R_only_SP_vs_tau_c_SSF}}
 		\subfloat[Achievable rate per UE for $M = 100$, $\tau_c = 200$ and $K =10$.]{
 			 			\begin{tikzpicture}
	\begin{semilogxaxis}[
	width=0.42\textwidth,
	height=0.32\textwidth,
	y label style = {at={(-0.02,0.5)},anchor=north},%,font=\small
	xlabel={Avg. SNR $\rho/\sigma^2$},
	ylabel={Avg. rate per UE [Mbit/s]},
	ymin = 10,
	ymax = 50,
	xmax = 1000,
	xmin= 0.1,
	ytick distance = 10,
	line width = 1.2pt,
	grid=both,
	%	 legend pos= north west,
	legend style={at={(0.8,0.5)},font=\footnotesize,line width=1pt,draw=black,mark size=0.1pt},%draw=none font=\fontsize{4}{5}\selectfont %(0.64,0.99)
	]% mark = none
	\addplot [only marks,mark=o, mark size=3pt,blue] table[x index={0}, y index={1}] {Figures/SP_only_vs_rho.txt};
	\addplot [solid,blue,solid] table[x index={0}, y index={2}] {Figures/SP_only_vs_rho.txt};
	\addplot [only marks,mark=triangle, mark size=4pt,green] table[x index={0}, y index={3},only marks] {Figures/SP_only_vs_rho.txt};
	\addplot [solid,green]  table[x index={0}, y index={4}] {Figures/SP_only_vs_rho.txt};
	\addplot [cyan,   only marks, mark=diamond, mark size=3pt] 	table[x index={0}, y index={5},only marks] {Figures/SP_only_vs_rho.txt};
	\addplot [magenta, densely dashed, mark=x, mark size=4pt, mark options={solid}] table[x index={0}, y index={6}] {Figures/SP_only_vs_rho.txt};
	\addplot [teal,  densely  dashed, mark=+, mark size=4pt, mark options={solid, line width = 1.6pt}] table[x index={0}, y index={7}] {Figures/SP_only_vs_rho.txt};
	\end{semilogxaxis}
% 		\draw (0.05,2.23) rectangle (3.66,4.1);
\end{tikzpicture}

% 'SP no sub. pilot'
% 'SP no sub. pilot'
% 'SP perf. sub. pilot'
% 'SP  perf. sub. pilot'
% 'SP est. sub. pilot'
% 'SP Approx. [25]'
% 'SP Approx. [27]'
 			 			\label{fig:R_only_SP_vs_rho_SSF}}
% 		\hfil 		
 	}\caption{ \color{black} Optimized parameters and SP achievable rates versus $M$, $\tau_c$ and $\rho/\sigma^2$. The solid lines correspond to the closed-form expressions in Theorem~\ref{th:ach_rate_ssf_SIP} and Corollary~\ref{cor:SP_SINR_UB}, the triangle and circle markers correspond to MC simulations over the SSF. All results are averaged over the LSF.}
 	\label{fig:ach_rates_SP_only_SSF_LSF}
 \end{figure*}
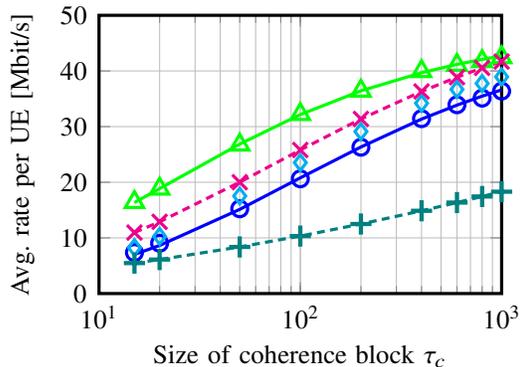
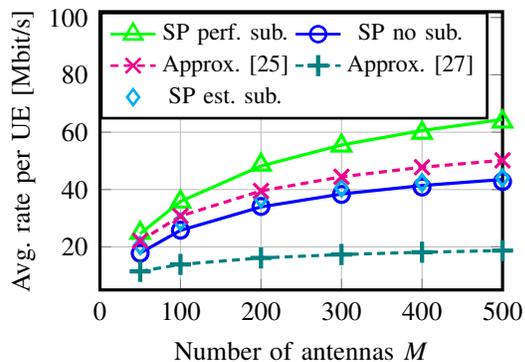
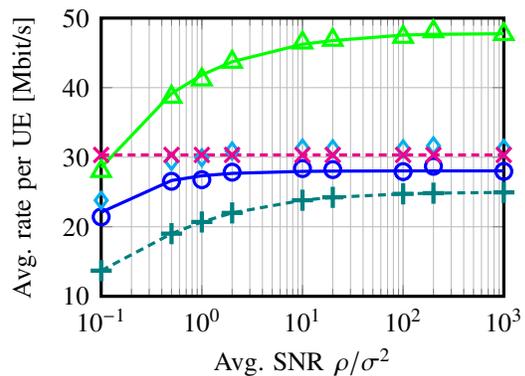
Fig.~\ref{tab:opt_param} shows a table with the average $\tau_p$ and $\Delta$ values that maximize the data rates. The optimal $\tau_p$ covers approximately 20\% of the coherence block and it increases with $M$ to counteract the effect of pilot contamination. With SP, the optimal $\Delta$ increases with the number of antennas; in line with the results from \cite{SIP_part1_KU_SA}. From the results with and without pilot subtraction, we can see that the optimal $\Delta$ seeks to balance the interference from pilot symbols and the quality of channel estimation. The rest of the graphs in Fig.~\ref{fig:ach_rates_SP_only_SSF_LSF} shows the average data rate per UE versus the number of antennas, size of the coherence block, and average SNR. The MC results confirm the validity of the closed-form expressions found in Theorem~\ref{th:ach_rate_ssf_SIP} and Corollary~\ref{cor:SP_SINR_UB}. It can be seen that there is a large gap between the results with no pilot subtraction and perfect pilot subtraction. However, the data rates with estimated pilot subtraction are closer to the data rates with no pilot subtraction, which is due to the cross products that arise from the correlation between channel estimates and data symbols. The approximation \cite{SIP_part1_KU_SA} is found within the results derived in Theorem~\ref{th:ach_rate_ssf_SIP} and Corollary~\ref{cor:SP_SINR_UB}, which makes it a good approximation of the performance. However, since the impact of noise is neglected in \cite{SIP_part1_KU_SA}, the approximation becomes less accurate in very low SNR scenarios. On the other hand, the approximation in \cite{VT_SP_approx_2016} greatly underestimates the performance with SP.}
\begin{figure*}[!t]
	\color{black}
	\centering
	{\captionsetup{width=0.4\textwidth}
		\subfloat[Achievable rate per UE for $\rho = \sigma^2/4$ (SNR~${=-6}$ dB), $\tau_c = 200$ and $K =10$.]{
				\begin{tikzpicture}
\usetikzlibrary{arrows}
	\begin{semilogxaxis}[
	width=0.42\textwidth,
	height=0.32\textwidth,
	y label style = {at={(-0.02,0.5)},anchor=north},
	xlabel={Number of antennas $M$},
	ylabel={Avg. rate per UE [Mbit/s]},
	extra x ticks={50},
	extra x tick labels={50},% \footnotesize
	xmin = 50,
	xmax = 100000,
	ymax = 120,
	ytick distance = 20,
	line width = 1.2pt,
	grid=both,
	%	 legend pos= north west,%(0.44,0.99)
	legend style={at={(1,0.61)},font=\footnotesize,line width=1pt,draw=black,mark size=0.1pt},%draw=none font=\fontsize{4}{5}\selectfont %(0.64,0.99)
%	legend entries={SP no sub. pilot, ,  SP perf. sub. pilot,  , SP Approx. \cite{SIP_part1_KU_SA}, SP Approx. \cite{VT_SP_approx_2016}},
%	legend entries={RP $\tau_p =K$, , ,
%					RP $\tau_p$ opt., , ,
%					SP no sub., , ,
%					SP perf. sub., , ,
%					SP est. sub.,
%					Approx. \cite{SIP_part1_KU_SA}, },
	]% mark = none
	\draw (rel axis cs:0.28,0.8) node[draw=none,fill =white,align = center] {\small{Asymptotic limits}};
	\draw[->,color = gray] (rel axis cs:0.1,0.85) -- (rel axis cs:0.2,0.92);	%lightgray
	\draw[->,color = gray] (rel axis cs:0.1,0.75) -- (rel axis cs:0.2,0.5);
	\draw[->,color = gray] (rel axis cs:0.08,0.75) -- (rel axis cs:0.05,0.3);
	\addplot [black,    only marks, mark=square,  mark size=2pt] 					   
		table[x index={0}, y index={1}] {Figures/R_vs_M_all_1e5.txt};							
	\addplot [black,  densely dashdotdotted  ] 									 					   
		table[x index={6}, y index={7}] {Figures/R_vs_M_all_1e5.txt};
	\addplot [black,  densely dotted  ] 									 					   
		table[x index={6}, y index={8}] {Figures/R_vs_M_all_1e5.txt};				
	\addplot [red,   only marks, mark=square*, mark size=2pt] 					   
		table[x index={0}, y index={2},only marks] {Figures/R_vs_M_all_1e5.txt};
	\addplot [red, loosely dashdotted]  									 					   
		table[x index={6}, y index={9}] {Figures/R_vs_M_all_1e5.txt};
	\addplot [red, densely dotted]  									 					   
		table[x index={6}, y index={10}] {Figures/R_vs_M_all_1e5.txt};				
	\addplot [blue,    only marks, mark=o,        mark size=3pt] 					   
		table[x index={0}, y index={3}] {Figures/R_vs_M_all_1e5.txt};
	\addplot [blue,    solid] 									 					   
		table[x index={6}, y index={11}] {Figures/R_vs_M_all_1e5.txt};
	\addplot [blue,   densely dotted] 									 					   
		table[x index={6}, y index={12}] {Figures/R_vs_M_all_1e5.txt};		
	\addplot [green,   only marks, mark=triangle, mark size=4pt] 					   
		table[x index={0}, y index={4},only marks] {Figures/R_vs_M_all_1e5.txt};
	\addplot [green,   solid]  									 					   
		table[x index={6}, y index={13}] {Figures/R_vs_M_all_1e5.txt};
	\addplot [green,   densely dotted]  									 					   
		table[x index={6}, y index={14}] {Figures/R_vs_M_all_1e5.txt};
	\addplot [cyan,   only marks, mark=diamond, mark size=3pt] 					   
	table[x index={0}, y index={5},only marks] {Figures/R_vs_M_all_1e5.txt};		
%	\addplot [olive,densely dashed, mark=x, mark size=4pt, mark options={solid}] 					   
%		table[x index={0}, y index={1}] {Figures/R_vs_M_all_1e5.txt};							
%	\addplot [olive,  densely dotted  ] 									 					   
%		table[x index={6}, y index={7}] {Figures/R_vs_M_all_1e5.txt};		
	\end{semilogxaxis}
%	\draw (0.05,1) rectangle (3.66,4.1);
%	\node [draw, text width=20pt] (4,4) {Asymptotic};
\end{tikzpicture}%
%  0   x
%  1   'RP'
%  2   'RP $\tau_p\,:$ optimized'
%  3   'SP no sub. pilot'
%  4   'SP perf. sub. pilot'
%  5   'SP est. sub. pilot'
%     
%  6     x
%  7   'RP '
%  8   'Asym. RP'
%  9   'RP $\tau_p\,:$ optimized'
%  10   'Asym. RP'
%  11  'SP no sub. pilot'
%  12   'Asym. SP  no sub. pilot'
%  13   'SP  perf. sub. pilot'
%  14   'Asym. SP  perf. sub. pilot'
%  15   'SP Approx. [25]'
%  16   'Asym. SP  Approx. [25]'

%%% original plot % R_vs_M
%\addplot [black,    only marks, mark=square,  mark size=2pt] 					   
%table[x index={0}, y index={1}] {Figures/R_vs_M.txt};
%\addplot [black,  densely dashdotdotted  ] 									 					   
%table[x index={0}, y index={2}] {Figures/R_vs_M.txt};
%\addplot [red,   only marks, mark=square*, mark size=2pt] 					   
%table[x index={0}, y index={3},only marks] {Figures/R_vs_M.txt};
%\addplot [red, loosely dashdotted]  									 					   
%table[x index={0}, y index={4}] {Figures/R_vs_M.txt};
%\addplot [blue,    only marks, mark=o,        mark size=3pt] 					   
%table[x index={0}, y index={5}] {Figures/R_vs_M.txt};
%\addplot [blue,    solid] 									 					   
%table[x index={0}, y index={6}] {Figures/R_vs_M.txt};
%\addplot [green,   only marks, mark=triangle, mark size=4pt] 					   
%table[x index={0}, y index={7},only marks] {Figures/R_vs_M.txt};
%\addplot [green,   solid]  									 					   
%table[x index={0}, y index={8}] {Figures/R_vs_M.txt};
%'RP'
%'RP '
%'RP $\tau_p\,:$ optimized'
%'RP $\tau_p\,:$ optimized'
%'SP no sub. pilot'
%'SP no sub. pilot'
%'SP perf. sub. pilot'
%'SP  perf. sub. pilot'
%'SP Approx. [25]'
			\label{fig:R_vs_M_SSF}}
		\subfloat[Achievable rate per UE for $\rho = \sigma^2/4$ (SNR~${=-6}$ dB), $M = 100$ and $K =10$.]{
				\begin{tikzpicture}
	\begin{semilogxaxis}[
	width=0.42\textwidth,
	height=0.32\textwidth,
	y label style = {at={(-0.02,0.5)},anchor=north},
	xlabel={Size of coherence block $\tau_c$},
	ylabel={Avg. rate per UE [Mbit/s]},
	ytick distance = 10,
	xmax = 1000,
	ymin = 5,
	ymax = 69,
	line width = 1.2pt,
	grid=both,
	legend columns = 3,
	transpose legend,
	%	 legend pos= north west,
	legend style={at={(0.92,1)},font=\footnotesize,line width=1pt,draw=black,mark size=0.1pt},%draw=none font=\fontsize{4}{5}\selectfont %(0.64,0.99)
%	legend entries={SP no sub. pilot, ,  SP perf. sub. pilot,  , SP Approx. \cite{SIP_part1_KU_SA}, SP Approx. \cite{VT_SP_approx_2016}},
%	legend entries={RP $\tau_p =K$, ,
%					RP $\tau_p$ opt., ,
%					SP no sub., ,
%					SP perf. sub.,  ,
%					SP est. sub.,
%					Approx. \cite{SIP_part1_KU_SA}},
	]% mark = none
	
	\addlegendentry{SP perf. sub.}
	\addlegendimage{solid,mark=triangle, mark size=4pt,green}
	
	\addlegendentry{RP $\tau_p$ opt.}	
	\addlegendimage{red, loosely dashdotted, mark=square*, mark size=2pt,mark options={solid}}
	
	\addlegendentry{SP est. sub.}
	\addlegendimage{only marks,mark=diamond, mark size=3pt,cyan}	
	
	\addlegendentry{SP no sub.}
	\addlegendimage{solid,mark=o, mark size=3pt,blue}
	
	\addlegendentry{RP $\tau_p =K$}	
	\addlegendimage{black, densely dashdotdotted , mark=square,  mark size=2pt,mark options={solid}}%,mark options={solid}
	
	\addlegendentry{}
	\addlegendimage{empty legend}	
	
	\addplot [black,    only marks, mark=square,  mark size=2pt] 					   
		table[x index={0}, y index={1}] {Figures/R_vs_tau_c.txt};
	\addplot [black,  densely dashdotdotted  ] 									 					   
		table[x index={0}, y index={2}] {Figures/R_vs_tau_c.txt};
	\addplot [red,   only marks, mark=square*, mark size=2pt] 					   
		table[x index={0}, y index={3},only marks] {Figures/R_vs_tau_c.txt};
	\addplot [red, loosely dashdotted]  									 					   
		table[x index={0}, y index={4}] {Figures/R_vs_tau_c.txt};
	\addplot [blue,    only marks, mark=o,        mark size=3pt] 					   
		table[x index={0}, y index={5}] {Figures/R_vs_tau_c.txt};
	\addplot [blue,    solid] 									 					   
		table[x index={0}, y index={6}] {Figures/R_vs_tau_c.txt};
	\addplot [green,   only marks, mark=triangle, mark size=4pt] 					   
		table[x index={0}, y index={7},only marks] {Figures/R_vs_tau_c.txt};
	\addplot [green,   solid]  									 					   
		table[x index={0}, y index={8}] {Figures/R_vs_tau_c.txt};
	\addplot [cyan,   only marks, mark=diamond, mark size=3pt] 					   
		table[x index={0}, y index={9},only marks] {Figures/R_vs_tau_c.txt};		
%	\addplot [olive,densely dashed, mark=x, mark size=4pt, mark options={solid}] 					   
%		table[x index={0}, y index={10}] {Figures/R_vs_tau_c.txt};							
	\end{semilogxaxis}
% 		\draw (0.05,2.23) rectangle (3.66,4.1);
\end{tikzpicture}
%     'RP'
%     'RP '
%     'RP $\tau_p\,:$ optimized'
%     'RP $\tau_p\,:$ optimized'
%     'SP no sub. pilot'
%     'SP no sub. pilot'
%     'SP perf. sub. pilot'
%     'SP  perf. sub. pilot'
%     'SP est. sub. pilot'
%     'SP Approx. [25]'

%	legend columns = 3,	
%	transpose legend,
%  mark options={solid, line width = 1.6pt}

%	
%	\addlegendentry{Approx. \cite{SIP_part1_KU_SA}}
%	\addlegendimage{densely dashed,mark=x, mark size=4pt, mark options={solid},magenta}
%	
			\label{fig:R_vs_S_SSF}}
		\\[-6pt]
		\subfloat[Achievable rate per UE for $M = 100$, $\tau_c = 200$ and $K =10$.]{
			\begin{tikzpicture}
	\begin{semilogxaxis}[
	width=0.42\textwidth,
	height=0.32\textwidth,
	y label style = {at={(-0.02,0.5)},anchor=north},
	xlabel={Avg. SNR $\rho/\sigma^2$},
	ylabel={Avg. rate per UE [Mbit/s]},
	ytick distance = 10,
	ymin = 10,
	ymax = 50,
	xmin = 0.1,
	xmax = 1000,
	line width = 1.2pt,
	grid=both,
	%	 legend pos= north west,
	legend style={at={(0.44,0.99)},font=\footnotesize,line width=1pt,draw=black,mark size=0.1pt},%draw=none font=\fontsize{4}{5}\selectfont %(0.64,0.99)
%	legend entries={SP no sub. pilot, ,  SP perf. sub. pilot,  , SP Approx. \cite{SIP_part1_KU_SA}, SP Approx. \cite{VT_SP_approx_2016}},
%	legend entries={RP $\tau_p =K$, ,
%					RP $\tau_p$ opt., ,
%					SP no sub., ,
%					SP perf. sub.,  ,
%					SP est. sub.,
%					Approx. \cite{SIP_part1_KU_SA}},
	]% mark = none
	\addplot [black,    only marks, mark=square,  mark size=2pt] 					   
		table[x index={0}, y index={1}] {Figures/R_vs_rho.txt};
	\addplot [black,  densely dashdotdotted  ] 									 					   
		table[x index={0}, y index={2}] {Figures/R_vs_rho.txt};
	\addplot [red,   only marks, mark=square*, mark size=2pt] 					   
		table[x index={0}, y index={3},only marks] {Figures/R_vs_rho.txt};
	\addplot [red, loosely dashdotted]  									 					   
		table[x index={0}, y index={4}] {Figures/R_vs_rho.txt};
	\addplot [blue,    only marks, mark=o,        mark size=3pt] 					   
		table[x index={0}, y index={5}] {Figures/R_vs_rho.txt};
	\addplot [blue,    solid] 									 					   
		table[x index={0}, y index={6}] {Figures/R_vs_rho.txt};
	\addplot [green,   only marks, mark=triangle, mark size=4pt] 					   
		table[x index={0}, y index={7},only marks] {Figures/R_vs_rho.txt};
	\addplot [green,   solid]  									 					   
		table[x index={0}, y index={8}] {Figures/R_vs_rho.txt};
	\addplot [cyan,   only marks, mark=diamond, mark size=3pt] 					   
		table[x index={0}, y index={9},only marks] {Figures/R_vs_rho.txt};		
%	\addplot [olive,densely dashed, mark=x, mark size=4pt, mark options={solid}] 					   
%		table[x index={0}, y index={10}] {Figures/R_vs_rho.txt};							
	\end{semilogxaxis}
% 		\draw (0.05,2.23) rectangle (3.66,4.1);
\end{tikzpicture}

%     'RP'
%     'RP '
%     'RP $\tau_p\,:$ optimized'
%     'RP $\tau_p\,:$ optimized'
%     'SP no sub. pilot'
%     'SP no sub. pilot'
%     'SP perf. sub. pilot'
%     'SP  perf. sub. pilot'
%     'SP est. sub. pilot'
%     'SP Approx. [25]'
			\label{fig:R_vs_rho_SSF}}
		\subfloat[Achievable sum rate per cell for $\rho = \sigma^2/4$ (SNR~${=-6}$ dB), $M = 100$ and $\tau_c =200$.]{
			\begin{tikzpicture}
	\begin{axis}[
	width=0.42\textwidth,
	height=0.32\textwidth,
	y label style = {align=center},%at={(-0.02,0.5)},anchor=north, 
	xlabel={Number of UEs per cell $K$},
	ylabel={\\[-20pt]Avg. sum rate\\[-8pt] per cell [Mbit/s]},
%	ytick distance = 10,
	xmin = 0,
	xmax = 150,
	line width = 1.2pt,
	grid=both,
	%	 legend pos= north west,
	legend style={at={(1,0.85)},font=\footnotesize,line width=1pt,draw=black,mark size=0.1pt},%draw=none font=\fontsize{4}{5}\selectfont %(0.64,0.99)
%	legend entries={SP no sub. pilot, ,  SP perf. sub. pilot,  , SP Approx. \cite{SIP_part1_KU_SA}, SP Approx. \cite{VT_SP_approx_2016}},
%	legend entries={RP $\tau_p =K$, ,
%					RP $\tau_p$ opt., ,
%					SP no sub., ,
%					SP perf. sub.,  ,
%					SP est.sub,
%					Approx. \cite{SIP_part1_KU_SA}},
	]% mark = none
	\addplot [black,    only marks, mark=square,  mark size=2pt] 					   
		table[x index={0}, y index={1}] {Figures/R_vs_K.txt};
	\addplot [black,  densely dashdotdotted  ] 									 					   
		table[x index={0}, y index={2}] {Figures/R_vs_K.txt};
	\addplot [red,   only marks, mark=square*, mark size=2pt] 					   
		table[x index={0}, y index={3},only marks] {Figures/R_vs_K.txt};
	\addplot [red, loosely dashdotted]  									 					   
		table[x index={0}, y index={4}] {Figures/R_vs_K.txt};
	\addplot [blue,    only marks, mark=o,        mark size=3pt] 					   
		table[x index={0}, y index={5}] {Figures/R_vs_K.txt};
	\addplot [blue,    solid] 									 					   
		table[x index={0}, y index={6}] {Figures/R_vs_K.txt};
	\addplot [green,   only marks, mark=triangle, mark size=4pt] 					   
		table[x index={0}, y index={7},only marks] {Figures/R_vs_K.txt};
	\addplot [green,   solid]  									 					   
		table[x index={0}, y index={8}] {Figures/R_vs_K.txt};
	\addplot [cyan,   only marks, mark=diamond, mark size=3pt] 					   
		table[x index={0}, y index={9},only marks] {Figures/R_vs_K.txt};		
%	\addplot [olive,densely dashed, mark=x, mark size=4pt, mark options={solid}] 					   
%		table[x index={0}, y index={10}] {Figures/R_vs_rho.txt};							
	\end{axis}
% 		\draw (0.05,2.23) rectangle (3.66,4.1);
\end{tikzpicture}
%     'RP'
%     'RP '
%     'RP $\tau_p\,:$ optimized'
%     'RP $\tau_p\,:$ optimized'
%     'SP no sub. pilot'
%     'SP no sub. pilot'
%     'SP perf. sub. pilot'
%     'SP  perf. sub. pilot'
%     'SP est. sub. pilot'
%     'SP Approx. [25]'
			\label{fig:R_vs_K_SSF}}
	}\caption{\color{black}Achievable rates versus $M$, $\tau_c$, $\rho/\sigma^2$ and $K$. The lines correspond to the closed-form expressions in Lemma~\ref{lem:ach_rate_ssf_RP}, Theorem~\ref{th:ach_rate_ssf_SIP} and Corollary~\ref{cor:SP_SINR_UB}. The markers correspond to MC simulations over the SSF. All results are averaged over the LSF.}
	\label{fig:ach_rates_SSF_LSF}
\end{figure*}
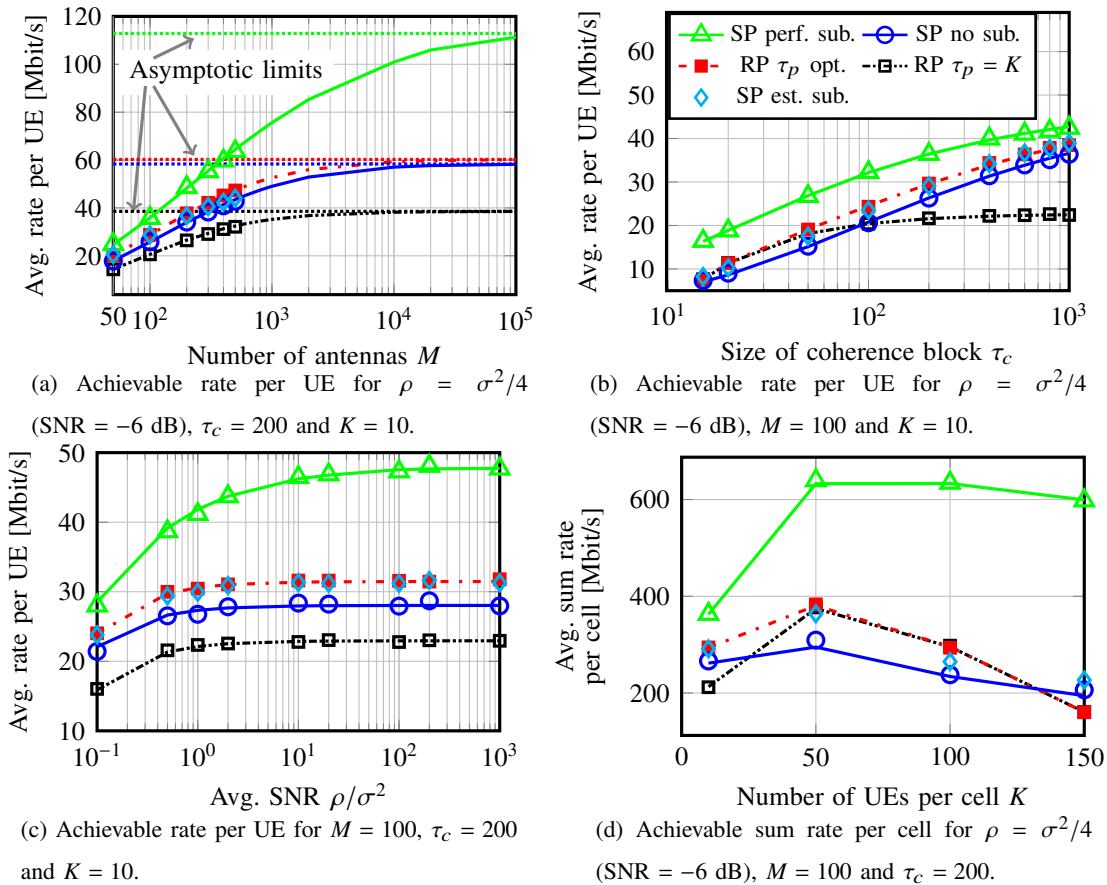
%\footnote{The results have been averaged over large-scale and SSF realizations.}
  
Fig.~\ref{fig:ach_rates_SSF_LSF} depicts the achievable rates per UE versus the number of BS antennas, coherence block size, and {\color{black}average SNR} along with the sum rate per cell versus the number UEs per cell. {\color{black}In general, we can see that SP outperforms RP with $\tau_p = K$ for most cases when $\tau_c$ is long enough to reduce pilot contamination. Otherwise, the data rates with RP, including the results when $\tau_p$ is optimized, provide comparable performance to that of SP with estimated pilot subtraction. When subtracting the pilot symbols perfectly, the data rates with SP provide the best performance, but it might be hard to achieve this in practice. In Fig.~\ref{fig:R_vs_M_SSF}, the asymptotic limits found derived in Corollary~\ref{cor:asymptotic_M_infty} are shown. We can see that more than $10^4$ BS antennas are needed to converge to the limits and the relative differences among the methods vary between the finite $M$ and $M\to \infty$.	}
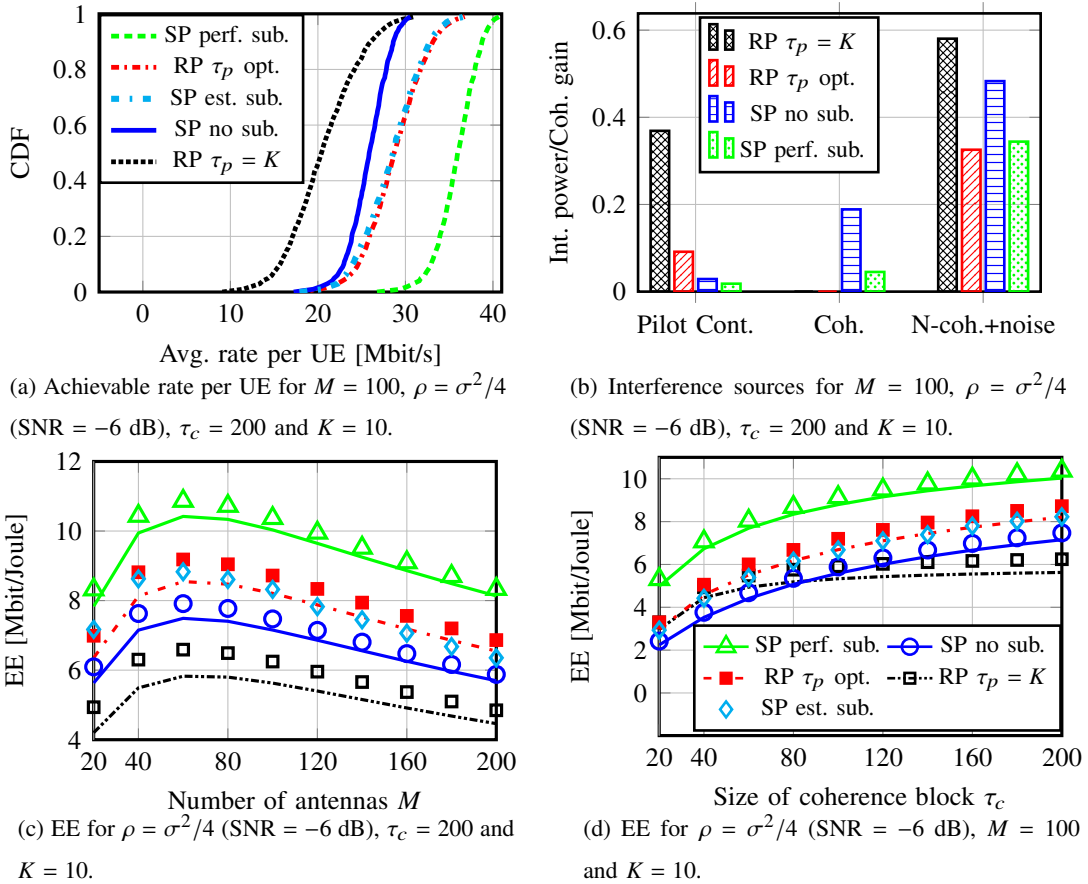
\begin{figure*}[!t]
	\color{black}
	\centering
	{\captionsetup{width=0.4\textwidth}		
		\subfloat[Achievable rate per UE for $M = 100$, $\rho = \sigma^2/4$ \mbox{(SNR $=-6$ dB)}, $\tau_c = 200$ and $K =10$.]{
			\begin{tikzpicture}
	\begin{axis}[
	width=0.42\textwidth,
	height=0.32\textwidth,
	y label style = {at={(-0.02,0.5)},anchor=north},
	xlabel={Avg. rate per UE [Mbit/s]},
	ylabel={CDF},
	xmax = 41,%16
	xmin = -5,
	ymin = 0,
	ymax = 1,
%	xtick distance = 4,
	line width = 1.2pt,
	grid=both,
	%	 legend pos= north west,
	legend style={at={(0.51,1)},font=\footnotesize,line width=1pt,draw=black,mark size=0.1pt},%draw=none font=\fontsize{4}{5}\selectfont %(0.64,0.99)
%	legend entries={SP no sub. pilot, ,  SP perf. sub. pilot,  , SP Approx. \cite{SIP_part1_KU_SA}, SP Approx. \cite{VT_SP_approx_2016}},
%	legend entries={RP $\tau_p =K$ ,
%					RP $\tau_p$ opt.,
%					SP no sub., 
%					SP perf. sub.,  
%					SP est. sub.,  
%					Approx. \cite{SIP_part1_KU_SA},},
	]% mark = none		
	
		\addlegendentry{SP perf. sub.}
		\addlegendimage{green,   densely dashed, line width = 1.8pt}
		
		\addlegendentry{RP $\tau_p$ opt.}	
		\addlegendimage{red,  dashdotted, line width = 1.8pt}
		
		\addlegendentry{SP est. sub.}
		\addlegendimage{cyan,   loosely dashdotted, line width = 2.3pt}	
		
		\addlegendentry{SP no sub.}
		\addlegendimage{blue,   solid, line width = 1.8pt}
		
		\addlegendentry{RP $\tau_p =K$}	
		\addlegendimage{black,  densely dotted, line width = 1.8pt}

	\addplot [black,  densely dotted, line width = 1.8pt  ] 									 					   
		table[x index={0}, y index={1}] {Figures/plot_cdf_all.txt};	
	\addplot [red,  dashdotted, line width = 1.8pt]  									 					   
		table[x index={2}, y index={3}] {Figures/plot_cdf_all.txt};
	\addplot [blue,   solid, line width = 1.8pt] 									 					   
		table[x index={4}, y index={5}] {Figures/plot_cdf_all.txt};	
	\addplot [green,   densely dashed, line width = 1.8pt]  									 					   
		table[x index={6}, y index={7}] {Figures/plot_cdf_all.txt};
	\addplot [cyan,   loosely dashdotted, line width = 2.3pt]  									 					   
		table[x index={8}, y index={9}] {Figures/plot_cdf_all.txt};
%	\addplot [green,   densely dotted]  									 					   
%		table[x index={8}, y index={9}] {Figures/plot_cdf_all.txt};
	\end{axis}
% 		\draw (0.05,2.23) rectangle (3.66,4.1);
\end{tikzpicture}

%'RP'
%'RP $\tau_p\,:$ optimized'
%'SP no sub. pilot'
%'SP  perf. sub. pilot'
%'SP Approx. [25]'
%plot_R_vs_M_inf
			\label{fig:R_cdf_lsf}}		
		\subfloat[Interference sources for $M = 100$, $\rho = \sigma^2/4$ \mbox{(SNR $=-6$ dB)}, $\tau_c =200$ and ${K=10}$.]{
			\begin{tikzpicture}
	\usetikzlibrary{patterns}
	\begin{axis}[
	width=0.42\textwidth,%0.42\textwidth
	height=0.32\textwidth,	
	ybar,
%	scale only axis,
	enlarge x limits=0.2,	
	y label style = {at={(-0.02,0.5)},anchor=north},%
	xlabel={{\color{white}$/$}},
	ylabel={Int. power/Coh. gain},	
	symbolic x coords = {Pilot Cont.,Coh.,N-coh.+noise},
	xtick = data,
%	x tick label style={rotate=-20,anchor=center},
%	nodes near coords,
	nodes near coords align={vertical},
	xticklabel style={text width=0.15\textwidth,align=center},%   x tick label style={rotate=45, anchor=east, align=center},
%	x tick label style={rotate=45,anchor=east},
%	xmax = 1000000,
%	xmin = 50,
	ymin = 0,
%	ymax = 100,
%	ytick distance = 20,
	bar width=7pt,
	line width = 1.0pt,
	grid=both,
	%	 legend pos= north west,
	legend style={at={(0.6,0.99)},font=\footnotesize,line width=1pt,draw=black,mark size=0.1pt},%draw=none font=\fontsize{4}{5}\selectfont %(0.64,0.99)
	legend entries={RP $\tau_p=K$,RP $\tau_p$ opt., SP no sub., SP perf. sub.},
	]% mark = none	 
	\addplot [black,pattern color = black, pattern = crosshatch] 
	coordinates	{(Pilot Cont.,3.6894604e-01) (Coh.,0.0000000e+00) (N-coh.+noise,5.8039305e-01)};	
	%table[x expr=\coordindex, y index={0}] {int_sources.txt};
	\addplot [red,pattern color = red, pattern = north east lines ]  									 					   
	coordinates	{(Pilot Cont.,9.1490195e-02) (Coh.,0.0000000e+00) (N-coh.+noise,3.2558640e-01)}; %table[x expr=\coordindex, y index={1}] {int_sources.txt};
	\addplot [blue,pattern color = blue,  pattern = horizontal lines] 									 					   
	coordinates	{(Pilot Cont.,2.8905140e-02) (Coh.,1.8873543e-01) (N-coh.+noise,4.8297304e-01)};%	table[x expr=\coordindex, y index={2}] {int_sources.txt};
	\addplot [green,pattern color = green, pattern = crosshatch dots]  									 					   
	coordinates	{(Pilot Cont.,1.8226205e-02) (Coh.,4.5128216e-02) (N-coh.+noise,3.4420148e-01)};%	table[x expr=\coordindex, y index={3}] {int_sources.txt};	
	\end{axis}
% 		\draw (0.05,2.23) rectangle (3.66,4.1);
\end{tikzpicture}

%            
%            
%            
%
%nodes near coords,
%nodes near coords align={vertical},

%'RP '
%'Asym. RP'
%'RP $\tau_p\,:$ optimized'
%'Asym. RP'
%'SP no sub. pilot'
%'Asym. SP  no sub. pilot'
%'SP  perf. sub. pilot'
%'Asym. SP  perf. sub. pilot'
%'SP Approx. [25]'
%'Asym. SP  Approx. [25]'
			\label{fig:int_term}}	
		\\[-7pt]
%		\subfloat[Achievable rate per UE for $M = 100$, $\rho = \sigma^2/4$ \mbox{(SNR $=-6$ dB)}, $\tau_c = 200$ and $K =10$.]{
%			\import{Figures/}{plot_cdf.tex}%plot_R_vs_M_inf
%			\label{fig:R_vs_M_inf}}		
		\subfloat[EE for $\rho = \sigma^2/4$ (SNR $=-6$ dB), $\tau_c = 200$ and $K =10$.]{
			\begin{tikzpicture}
	\begin{axis}[
	width=0.42\textwidth,
	height=0.32\textwidth,
	y label style = {at={(-0.02,0.5)},anchor=north},
	xlabel={Number of antennas $M$},
	ylabel={EE [Mbit/Joule]},
	extra x ticks={20},
	extra x tick labels={20},% \footnotesize
	ymin = 4,
	ymax = 12,
	xmin = 20,
	xmax = 200,	
	xtick distance = 40,
	line width = 1.2pt,
	grid=both,
	%	 legend pos= north west,
	legend style={at={(0.99,0.99)},font=\footnotesize,line width=1pt,draw=black,mark size=0.1pt},%draw=none font=\fontsize{4}{5}\selectfont %(0.64,0.99)
%	legend entries={SP no sub. pilot, ,  SP perf. sub. pilot,  , SP Approx. \cite{SIP_part1_KU_SA}, SP Approx. \cite{VT_SP_approx_2016}},
%	legend entries={RP $\tau_p =K$, ,
%					RP $\tau_p$ opt., ,
%					SP no sub., ,
%					SP perf. sub.,  ,
%					SP est. sub.,
%					Approx. \cite{SIP_part1_KU_SA}},				
	]% mark = none
	\addplot [black,    only marks, mark=square,  mark size=2pt] 					   
		table[x index={0}, y index={1}] {Figures/EE_vs_M.txt};
	\addplot [black,  densely dashdotdotted  ] 									 					   
		table[x index={0}, y index={2}] {Figures/EE_vs_M.txt};
	\addplot [red,   only marks, mark=square*, mark size=2pt] 					   
		table[x index={0}, y index={3},only marks] {Figures/EE_vs_M.txt};
	\addplot [red, loosely dashdotted]  									 					   
		table[x index={0}, y index={4}] {Figures/EE_vs_M.txt};
	\addplot [blue,    only marks, mark=o,        mark size=3pt] 					   
		table[x index={0}, y index={5}] {Figures/EE_vs_M.txt};
	\addplot [blue,    solid] 									 					   
		table[x index={0}, y index={6}] {Figures/EE_vs_M.txt};
	\addplot [green,   only marks, mark=triangle, mark size=4pt] 					   
		table[x index={0}, y index={7},only marks] {Figures/EE_vs_M.txt};
	\addplot [green,   solid]  									 					   
		table[x index={0}, y index={8}] {Figures/EE_vs_M.txt};
	\addplot [cyan,   only marks, mark=diamond, mark size=3pt] 					   
		table[x index={0}, y index={9},only marks] {Figures/EE_vs_M.txt};		
%	\addplot [olive,densely dashed, mark=x, mark size=4pt, mark options={solid}] 					   
%		table[x index={0}, y index={10}] {Figures/EE_vs_M.txt};							
	\end{axis}
% 		\draw (0.05,2.23) rectangle (3.66,4.1);
\end{tikzpicture}
%'RP '
%'L.B. RP'
%'RP $\tau_p\,:$ optimized'
%'L.B. RP'
%'SP no sub. pilot'
%'SP no sub. pilot L.B.'
%'SP  perf. sub. pilot'
%'L.B. SP  perf. sub. pilot'
%'SP est. sub. pilot'
%'SP Approx. [25]'
			\label{fig:EE_vs_M}}			
		\subfloat[EE for $\rho = \sigma^2/4$ (SNR $=-6$ dB), $M = 100$ and $K =10$.]{
			\begin{tikzpicture}
	\begin{axis}[
	width=0.42\textwidth,
	height=0.32\textwidth,
	y label style = {at={(-0.02,0.5)},anchor=north},
	xlabel={Size of coherence block $\tau_c$},
	ylabel={EE [Mbit/Joule]},
	extra x ticks={20},
	extra x tick labels={20},% \footnotesize	
	ymin = -1.99,
	ymax = 11,
	xmin = 20,
	xmax = 200,
	xtick distance = 40,	
	ytick distance = 2,	
	line width = 1.2pt,
	grid=both,
	%	 legend pos= north west,
	legend columns = 3,
	transpose legend,
	legend style={at={(1,0.4)},font=\footnotesize,line width=1pt,draw=black,mark size=0.1pt},%draw=none font=\fontsize{4}{5}\selectfont %(0.64,0.99)
%	legend entries={SP no sub. pilot, ,  SP perf. sub. pilot,  , SP Approx. \cite{SIP_part1_KU_SA}, SP Approx. \cite{VT_SP_approx_2016}},
%	legend entries={RP $\tau_p =K$, ,
%					RP $\tau_p$ opt., ,
%					SP no sub., ,
%					SP perf. sub.,  ,
%					SP est. sub.,
%					Approx. \cite{SIP_part1_KU_SA}},				
	]% mark = none
		\addlegendentry{SP perf. sub.}
		\addlegendimage{solid,mark=triangle, mark size=4pt,green}
		
		\addlegendentry{RP $\tau_p$ opt.}	
		\addlegendimage{red, loosely dashdotted, mark=square*, mark size=2pt,mark options={solid}}
		
		\addlegendentry{SP est. sub.}
		\addlegendimage{only marks,mark=diamond, mark size=3pt,cyan}	
		
		\addlegendentry{SP no sub.}
		\addlegendimage{solid,mark=o, mark size=3pt,blue}
		
		\addlegendentry{RP $\tau_p =K$}	
		\addlegendimage{black, densely dashdotdotted , mark=square,  mark size=2pt,mark options={solid}}%,mark options={solid}
		
		\addlegendentry{}
		\addlegendimage{empty legend}	
		
	\addplot [black,    only marks, mark=square,  mark size=2pt] 					   
		table[x index={0}, y index={1}] {Figures/EE_vs_tau_c.txt};
	\addplot [black,  densely dashdotdotted  ] 									 					   
		table[x index={0}, y index={2}] {Figures/EE_vs_tau_c.txt};
	\addplot [red,   only marks, mark=square*, mark size=2pt] 					   
		table[x index={0}, y index={3},only marks] {Figures/EE_vs_tau_c.txt};
	\addplot [red, loosely dashdotted]  									 					   
		table[x index={0}, y index={4}] {Figures/EE_vs_tau_c.txt};
	\addplot [blue,    only marks, mark=o,        mark size=3pt] 					   
		table[x index={0}, y index={5}] {Figures/EE_vs_tau_c.txt};
	\addplot [blue,    solid] 									 					   
		table[x index={0}, y index={6}] {Figures/EE_vs_tau_c.txt};
	\addplot [green,   only marks, mark=triangle, mark size=4pt] 					   
		table[x index={0}, y index={7},only marks] {Figures/EE_vs_tau_c.txt};
	\addplot [green,   solid]  									 					   
		table[x index={0}, y index={8}] {Figures/EE_vs_tau_c.txt};
	\addplot [cyan,   only marks, mark=diamond, mark size=3pt] 					   
		table[x index={0}, y index={9},only marks] {Figures/EE_vs_tau_c.txt};		
	%	\addplot [olive,densely dashed, mark=x, mark size=4pt, mark options={solid}] 					   
	%		table[x index={0}, y index={10}] {Figures/EE_vs_tau_c.txt};							
	\end{axis}
% 		\draw (0.05,2.23) rectangle (3.66,4.1);
\end{tikzpicture}
%'RP '
%'L.B. RP'
%'RP $\tau_p\,:$ optimized'
%'L.B. RP'
%'SP no sub. pilot'
%'SP no sub. pilot L.B.'
%'SP  perf. sub. pilot'
%'L.B. SP  perf. sub. pilot'
%'SP est. sub. pilot'
%'SP Approx. [25]'
			\label{fig:EE_vs_S}}	
	}\caption{\color{black}Achievable rates CDF, interference sources received power over coherent gain, and EE versus $M$ and $\tau_c$. In Figs.~\ref{fig:EE_vs_M} and~\ref{fig:EE_vs_S} the markers are based on the closed-form expressions in Lemma~\ref{lem:ach_rate_ssf_RP}, Theorem~\ref{th:ach_rate_ssf_SIP} and Corollary~\ref{cor:SP_SINR_UB}, averaged over the LSF. The lines are based on the closed-form expressions in Theorem~\ref{th:LB_ach_rate_wrt_d}.}
	%\label{fig_sim}
\end{figure*}

{\color{black}Fig.~\ref{fig:R_cdf_lsf} depicts the cumulative distribution function (CDF) of the achievable rates per UE for different realizations of LSF. We can see that the rate distribution does not show any large difference between the different methods. Fig.~\ref{fig:int_term} shows the strength of the interference sources with respect to the coherent gain (all terms are defined in Table~\ref{tab:ach_rate_comp}). We see that with SP, there is a reduction of the pilot contamination contributions with respect to RP. At the same time, however, additional coherent interference appears from data transmission and, in the case of SP without pilot subtraction, that is substantial. It is important to note that the overall impact of coherent interference for $M =100$ and $K=10$ is lower than the impact of non-coherent interference with both RP and SP. This suggests that, in practical dense deployments, Massive MIMO systems may not be limited by coherent interference. Fig.~\ref{fig:EE_vs_M} and Fig.~\ref{fig:EE_vs_S} depict the EE in terms of the number of BS antennas and size of the coherence block respectively. We see that the closed-form lower bounds found in Theorem~\ref{th:LB_ach_rate_wrt_d} follow the same trend as the MC simulations. In general, we can see a similar trend as in previous results, SP outperforms RP when $\tau_p=K$, however when we optimize $\tau_p$ we find that RP provides comparable EE than SP with estimated pilot subtraction. The highest EE is found with SP when pilot symbols are subtracted perfectly. }

\section{Conclusions}
\label{sec:conclusion}
{\color{black}In this paper, we derived the first rigorous achievable rate expression for a multicell Massive MIMO network with SP. We analytically and numerically compared RP and SP in a practical multicell deployment.} By examining the contribution of different sources of interference, we observed that SP is able to reduce pilot contamination at the expense of incorporating further coherent and non-coherent interference that limits the system performance. {\color{black}The results showed that, by optimizing the pilot length with RP, the average SE and EE are comparable to SP when estimated pilot subtraction is used. On the other hand, when the pilot symbols are subtracted perfectly with SP, the SE and EE are the highest, which indicates that there is room for improvement---iterative decoding algorithms might be able to bridge this gap. When analyzing the large number of BS antennas regime, we encountered that asymptotic results do not convey accurate results to gain insights into the behavior of practical deployments. Moreover, it is worth to stress that in practical deployments the effect of coherent interference, such as pilot contamination, can be less influential than non-coherent inter-cell on the SE of UEs. 

The use of SP has the potential to provide better performance by using other signal processing schemes like zero-forcing, multicell MMSE decoding, or iterative decoding algorithms. However, it is not clear whether the benefit of using such schemes would compensate for the increased computational complexity. All this study is left for future work.}

\appendices

\section{Proof of Theorem~\ref{th:ach_rate_ssf_SIP}}
\label{app:th_1}
Since the channels are circularly symmetric complex Gaussian random vectors, the channel gain uncertainty term (i.e., the first term in the denominator in \eqref{eq:SINR_SIP_0k}) can be computed as (e.g., \cite[Appendix A]{T_Marzetta_Ma_MIMO_book})
\begin{align}
\frac{\E\left\{  \left\|\h_{00k}\right\|^4\right\} -\left|\E\left\{ \left\|\h_{00k}\right\|^2 \right\}\right|^2}{M \beta_{00k}}  &= \frac{M\left(M +  1\right)\beta_{00k}^2 -M^2 \beta_{00k}^2 }{M \beta_{00k}} =  \beta_{00k}.
\label{eq:ch_gain_unc_cf}
\end{align}
To calculate  the variance of the rest of the effective noise, we first condition on an arbitrary realization of $\rchi_{l'i}^\textsc{sp}$ $\forall l' \in \Phi_D$, $i\in \{1,\ldots,K\}$ and then compute the expectation over $\rchi_{l'i}^\textsc{sp}$ as
\begin{align}
\Var(n_{\textit{eff}}) = \E\left\{ \left|n_\textit{eff} - \E\{n_\textit{eff}\}\right|^2\right\} = \E\left\{\E\left\{\left. \left|n_\textit{eff}\right|^2\right| \rchi_{l'i}^\textsc{sp}\right\}\right\} - \left|\E\left\{\E\left\{\left.n_\textit{eff}\right| \rchi_{l'i}^\textsc{sp}\right\}\right\}\right|^2. 
\label{eq:Var_n_eff_SP}
\end{align}
To proceed further, the following lemma is needed.
\begin{lemma}(\cite[Appendix A]{T_Marzetta_Ma_MIMO_book})
	Consider two independent random vectors distributed as ${\x\sim\CN\left(\mathbf{0},\sigma^2_x \I_M\right)}$ and ${\y\sim\CN\left(\mathbf{0},\sigma^2_y \I_M\right)}$, then we have the following results:
	\begin{align}
	\label{eq:E_second_moment}
	&\begin{aligned}
	\E\left\{\left(\x +\y \right)^H \x \right\} &= \E\left\{\left\| \x\right\|^2 \right\} = M \sigma_x^2
	\end{aligned}\\
	\label{eq:E_fourth_moment}
	&\begin{aligned}
	\E\left\{\left|\left(\x +\y \right)^H \x\right|^2 \right\} 
	& =  M(M + 1) \sigma_x^4 + M \sigma_x^2 \sigma_y^2.
	\end{aligned}
	\end{align}
	\label{lem:moments}
\end{lemma}
%\begin{IEEEproof}
%	It follows from straightforward calculations, e.g. \cite[Appendix A]{T_Marzetta_Ma_MIMO_book}.
%\end{IEEEproof}
By applying Lemma~\ref{lem:moments} {\color{black} and the result from \eqref{eq:rnd_pilot_SP}} to \eqref{eq:Var_n_eff_SP}, we have that 
\begin{align*}
&\E\left\{\E\left\{\left.n_\textit{eff}\right| \rchi_{l'i}^\textsc{sp}\right\}\right\}\\
&= \sqrt{\frac{M}{q_{0k} \beta_{00k}}}\E\left\{ \left( \frac{p_{0k} \beta_{00k}}{\tau_c}[\bphi_{0k}]_j
+ \sum_{l'\in \Phi_{D}} \sum_{i=1}^{K} \left( \rchi_{l'i}^\textsc{sp} q_{l'i}[\bphi_{l'i}]_j +  \xi_{l'i}\frac{p_{l'i}}{\tau_c} [\bphi_{0k}]_j\right)\beta_{0l'i}   + \frac{\sigma^2 [\bphi_{0k}]_j}{\tau_c}\right)\right\}\\
&= {\color{black}\sqrt{\frac{M}{q_{0k} \beta_{00k}}} \left( q_{0k}\beta_{00k} 
+ \frac{1}{\tau_c}\sum_{l'\in \Psi_{D}}\sum_{i=1}^{K}  q_{l'i} \beta_{0l'i} + \frac{1}{\tau_c} \sum_{l'\in \Phi_{D}} \sum_{i=1}^{K} p_{l'i}\beta_{0l'i}   + \frac{\sigma^2 }{\tau_c}\right)[\bphi_{0k}]_j}
\stepcounter{equation}\tag{\theequation}\label{eq:mean_neff}
\end{align*}
and
\begin{align*}\allowdisplaybreaks
&\E\left\{\E\left\{\left. \left|n_\textit{eff}\right|^2\right| \rchi_{l'i}^\textsc{sp}\right\}\right\}\\
 & = \E\Bigg\{\E\left\{\left.p_{0k}\left|\bar{\v}_{00k}^H\h_{00k}[\s_{0k}]_j \right|^2+\left|\v_{00k}^H\n_{0j}\right|^2 + \left|	\sum_{l'\in\Phi_{D}}\sum_{i= 1}^K\left( \sqrt{q_{l'i}}[\bphi_{l'i}]_j +\xi_{l'i}  \sqrt{p_{l'i}} [\s_{l'i}]_j\right)\v_{00k}^H\h_{0l'i}\right|^2 \right| \rchi_{l'i}^\textsc{sp}\right\}\\
 % corr 3 1
 &\mkern 30mu + 2\:\Re\left( \E\left\{\v_{00k}^H\n_{0j} \sqrt{p_{0k}}\h_{00k}^H \bar{\v}_{00k}[\s_{0k}]_j^*\Big| \rchi_{l'i}^\textsc{sp}\right\}  \right)\\
 % corr 3 2
 &\mkern 30mu + 2\:\Re\left( \E\left\{\v_{00k}^H\n_{0j} \sum_{l'\in\Phi_{D}}\sum_{i= 1}^K\h_{0l'i}^H\v_{00k} \left( \sqrt{q_{l'i}}[\bphi_{l'i}]_j^* + \xi_{l'i}\sqrt{p_{l'i}} [\s_{l'i}]_j^*\right)\Bigg| \rchi_{l'i}^\textsc{sp} \right\}  \right)\\%\displaybreak
 % corr 1 2
 &\mkern 30mu + 2\:\Re\left( \E\left\{\sqrt{p_{0k}} \bar{\v}_{00k}^H\h_{00k}[\s_{0k}]_j\sum_{l'\in\Phi_{D}}\sum_{i= 1}^K\h_{0l'i}^H\v_{00k} \left( \sqrt{q_{l'i}}[\bphi_{l'i}]_j^* +\xi_{l'i} \sqrt{p_{l'i}} [\s_{l'i}]_j^*\right)\Bigg| \rchi_{l'i}^\textsc{sp}
 \right\}  \right)\Bigg\}
 % *************** next equality ****************************
\\
&= p_{0k}\beta_{00k}\left(\frac{1}{\gamma_{0k}^\textsc{sp}}  - 1\right)  + \frac{M}{\tau_c}  \frac{p_{0k}}{q_{0k}} p_{0k} \beta_{00k} + \frac{\left(M + 1\right)}{\tau_c^2}  \frac{p_{0k}}{q_{0k}} p_{0k} \beta_{00k}
+  \frac{\sigma^2}{\gamma_{0k}^\textsc{sp}} + M\frac{ \sigma^4}{\tau_c^2 q_{0k}\beta_{00k} }
\\
&{\color{black}
\mkern 10mu + 
\frac{M}{q_{0k}\beta_{00k}} \left(\left(q_{0k}\beta_{00k} + \frac{1}{\tau_c}\sum_{l'\in\Psi_{D}}\sum_{i=1}^{K}q_{l'i} \beta_{0l'i}  +  \sum_{l'\in \Phi_{D}}\sum_{i=1}^{K} \xi_{l'i}\frac{p_{l'i}}{\tau_c} \beta_{0l'i} \right)^2 + \frac{1}{\tau_c}\left(1 - \frac{1}{\tau_c}\right)\sum_{l'\in\Psi_{D}}\sum_{i=1}^{K}q_{l'i}^2 \beta_{0l'i}^2\right) }\\ % 
&{\color{black}
\mkern 10mu +
\frac{M}{q_{0k}\beta_{00k}} \left( \frac{1}{\tau_c}\sum_{l'\in\Psi_{D}}\sum_{i=1}^{K}p_{l'i}q_{l'i}\beta_{0l'i}^2 + \frac{1}{\tau_c}\sum_{l'\in \Phi_{D}}\sum_{i =1}^{K}\left(q_{l'i} + \xi_{l'i}p_{l'i} \right)p_{l'i}\beta_{0l'i}^2 \right) }
\\
&{\color{black}
\mkern 10mu  +
\frac{1}{\tau_c q_{0k}\beta_{00k}} \left(\frac{2}{\tau_c}\sum_{l'\in\Psi_{D}}\sum_{i=1}^{K}p_{l'i}q_{l'i}\beta_{0l'i}^2   + \frac{1}{\tau_c} \sum_{l'\in \Phi_{D}}\sum_{i=1}^{K}\xi_{l'i} p_{l'i}^2\beta_{0l'i}^2 \right)
 +
\frac{1}{\gamma_{0k}^\textsc{sp}}\left( \sum_{l'\in\Phi_{D}}\sum_{i=1}^{K}\left(q_{l'i} +\xi_{l'i} p_{l'i}\right) \beta_{0l'i} \right)
}\\
&{\color{black}
\mkern 10mu 
+ 2  \frac{M \sigma^2}{\tau_c^2} \frac{p_{0k}}{q_{0k}}+ 2 \frac{M \sigma^2}{\tau_c q_{0k}\beta_{00k}}\left(  q_{0k}\beta_{00k} + \frac{1}{\tau_c} \sum_{l'\in\Psi_{D}}\sum_{i=1}^{K}  q_{l'i} \beta_{0l'i} + \sum_{l'\in \Phi_{D}}\sum_{i=1}^{K}\xi_{l'i} \frac{p_{l'i}}{\tau_c} \beta_{0l'i}    \right)
}\\
&{\color{black}
\mkern 10mu +
2\left(    \frac{p_{0k}\beta_{00k}}{\tau_c} +  \frac{M}{\tau_c} \frac{p_{0k}}{q_{0k}} \left( q_{0k}\beta_{00k}+   \frac{1}{\tau_c}\sum_{l'\in\Psi_{D}}\sum_{i=1}^{K}q_{l'i}\beta_{0l'i} + \sum_{l'\in \Phi_{D}}\sum_{i=1}^{K}\xi_{l'i} \frac{p_{l'i}}{\tau_c}\beta_{0l'i}  \right)   \right) .}
\stepcounter{equation}\tag{\theequation}\label{eq:fourth_moment_neff}
\end{align*}%\setcounter{equation}{56}
By combining \eqref{eq:mean_neff} with \eqref{eq:fourth_moment_neff}, we have that 
{\color{black}\begin{align*}\allowdisplaybreaks
\Var(n_{\textit{eff}}) = & 
\frac{M}{q_{0k}\beta_{00k}} \left( \frac{1}{\tau_c}\sum_{l'\in\Psi_{D}}\sum_{i=1}^{K}\left(p_{l'i} + \left(1 - \frac{1}{\tau_c}\right)q_{l'i}\right)q_{l'i}\beta_{0l'i}^2 + \frac{1}{\tau_c}\sum_{l'\in \Phi_{D}}\sum_{i =1}^{K}\left(q_{l'i} + p_{l'i} \right)p_{l'i}\beta_{0l'i}^2 \right)
\\
&\mkern 0mu  +
\frac{1}{\tau_c q_{0k}\beta_{00k}} \left(\frac{2}{\tau_c}\sum_{l'\in\Psi_{D}}\sum_{i=1}^{K}p_{l'i}q_{l'i}\beta_{0l'i}^2   + \frac{1}{\tau_c} \sum_{l'\in \Phi_{D}}\sum_{i=1}^{K} p_{l'i}^2\beta_{0l'i}^2 \right)
\\
&\mkern 0mu
-p_{0k}\beta_{00k} +  \frac{2}{\tau_c} p_{0k} \beta_{00k}+
\frac{1}{\gamma_{0k}^\textsc{sp}}\left( \sum_{l'\in\Phi_{D}}\sum_{i=1}^{K}\left(q_{l'i} + p_{l'i}\right) \beta_{0l'i} + \sigma^2 \right)
\stepcounter{equation}\tag{\theequation}\label{eq:var_neff_final}
\end{align*}}%\setcounter{equation}{57}
and by combining \eqref{eq:SINR_SIP_0k} and \eqref{eq:gamma_sip_gen} with \eqref{eq:ch_gain_unc_cf} and \eqref{eq:var_neff_final}, the proof is complete.
\section{Proof of Corollary~\ref{cor:realtion_rate_RP_SP}}
\label{app:relation_rate_RP_SP}
By studying the first and second derivative of $\AchR_{0k}^{\textsc{rp-a}}$ with respect to $\zeta$, we have
\begin{align}
\frac{\partial \AchR_{0k}^{\textsc{rp-a}}(\zeta)}{\partial \zeta} &= \left(-\frac{\ln\left(1 + \zeta \SIR_\textsc{rp}\right)}{\ln(2)} + \frac{\left(1 - \zeta\right)\SIR_\textsc{rp}}{\ln(2)\left(1 + \zeta \SIR_\textsc{rp}\right)}\right)\Bw\\
\frac{\partial^2 \AchR_{0k}^{\textsc{rp-a}}(\zeta)}{\partial \zeta^2} &= - \frac{\left(2 + \left(1 + \zeta\right)\SIR_\textsc{rp}\right) }{ \SIR_\textsc{rp}\ln(2) \left(1 + \zeta \SIR_\textsc{rp}\right)^2} \Bw < 0\;. 
\end{align}
We can see that $\AchR_{0k}^{\textsc{rp-a}}(\zeta)$ is a concave function and $\AchR_{0k}^{\textsc{rp-a}}(0) = \AchR_{0k}^{\textsc{rp-a}}(1) = 0$. Thus, considering that $\SIR_\textsc{rp}>0$, the maximum point of $\AchR_{0k}^{\textsc{rp-a}}(\zeta)$ is obtained when its derivative is zero and it is found at $\zeta^{\texttt{max}}$ shown in \eqref{eq:zeta_min}. This concludes the proof.

\section{Proof of Theorem~\ref{th:LB_ach_rate_wrt_d}}
\label{app:proof_th_LB_ach_rate_wrt_d}

{\color{black}By introducing the definitions of $q_{l'i} = p_{l'i} =\rho /\beta_{l'l'i}$ with RP, $q_{l'i}=\Delta\rho /\beta_{l'l'i}$ and ${p_{l'i}=(1 -\Delta)\rho /\beta_{l'l'i}}$ with SP and, $\beta_{l'l'i}= \omega^{-1} d_{l'l'i}^{-\alpha}$, into \eqref{eq:SINR_rp_gen}, \eqref{eq:SINR_sip_gen} and \eqref{eq:SINR_sip_gen_perf_pilot_sub} we have
\begin{align*}
&\SINR_{0k}^\textsc{rp} = \frac{M}{ \frac{M}{\tau_p} \sum\limits_{l'\in\Psi_{D}} \!\sum\limits_{i=1}^{K}\!\left(\frac{d_{l'l'i}^{\alpha}}{d_{0l'i}^{\alpha}} \right)^2  +   \left(1 +  \frac{1}{\tau_p}\sum\limits_{l'\in\Psi_{D}}\!\sum\limits_{i=1}^{K} \frac{d_{l'l'i}^{\alpha}}{d_{0l'i}^{\alpha}}  + \frac{\sigma^2}{\rho \tau_p}\right)\left(K  +     \sum\limits_{l'\in\Psi_{D}} \sum\limits_{i=1}^{K}\frac{d_{l'l'i}^{\alpha}}{d_{0l'i}^{\alpha}}  + \frac{\sigma^2}{\rho}\right)}
	\stepcounter{equation}\tag{\theequation}\label{eq:SINR_ch_inv_RP}
	\\
&
\SINR_{0k}^\textsc{sp} =M (1 - \Delta)\Bigg/\left(\frac{M}{\tau_c}\!\left(\!1 \!-\! \frac{\Delta}{\tau_c}\!\right)\mkern -7mu\sum_{l'\in \Psi_{D}}\!\sum_{i =1}^{K}\! \left(\frac{d_{l'l'i}^\alpha}{d_{0l'i}^\alpha}\right)^{\!2} \!+ \frac{M}{\tau_c}\frac{(1 - \Delta)}{\Delta}\left(\!K\!+\!\!\!\sum_{l'\in \Psi_{D}}\!\sum_{i =1}^{K}\left(\frac{d_{l'l'i}^\alpha}{d_{0l'i}^\alpha}\right)^{\!2}\right) 
\right.\\
&\mkern 60mu	
 +\frac{2}{\tau_c}(1 - \Delta)+\frac{2(1 - \Delta)}{\tau_c^2}\!\! \sum_{l'\in \Psi_{D}}\!\sum_{i =1}^{K}\!\left(\frac{d_{l'l'i}^\alpha}{d_{0l'i}^\alpha}\right)^{\!2}
+   \frac{(1 - \Delta)^2}{\tau_c^2\Delta }\left(K\!+\!\!\!\!\sum_{l'\in \Psi_{D}}\sum_{i=1}^{K}\left(\frac{d_{l'l'i}^\alpha}{d_{0l'i}^\alpha}\right)^{\!2}\right) 
\\
&\mkern 60mu		
\left.
+\!\left(\! 1 + \frac{1}{\Delta \tau_c}\sum\limits_{l'\in\Psi_D}\!\sum\limits_{i =1}^K \frac{d_{l'l'i}^\alpha}{d_{0l'i}^\alpha} \!+\frac{K(1 - \Delta)}{\Delta\tau_c } \! +\! \frac{\sigma^2}{\Delta\rho\tau_c}\right)\!\left( \!K\!+\!\!\!\sum_{l'\in\Psi_{D}}\!\sum_{i=1}^{K}\! \frac{d_{l'l'i}^\alpha}{d_{0l'i}^\alpha} \!+\!\frac{\sigma^2}{\rho}\!\right)
\right)	\stepcounter{equation}\tag{\theequation}\label{eq:SINR_ch_inv_SP}
\\
&\leq
\SINR_{0k}^\textsc{sp-ub} = M (1 - \Delta)\Bigg/\left(\frac{M(1 -\Delta)}{\tau_c}\!\mkern -5mu\sum_{l'\in \Psi_{D}}\!\sum_{i =1}^{K}\! \left(\frac{d_{l'l'i}^\alpha}{d_{0l'i}^\alpha}\right)^{\!2} \!+ \frac{M}{\tau_c}\frac{(1 - \Delta)^2}{\Delta}\left(\!K\!+\!\!\!\sum_{l'\in \Psi_{D}}\!\sum_{i =1}^{K}\left(\frac{d_{l'l'i}^\alpha}{d_{0l'i}^\alpha}\right)^{\!2}\right) 
\right.
\\
&\mkern 100mu	
+   \frac{(1 - \Delta)^2}{\tau_c^2\Delta }\left(K\!+\!\!\!\!\sum_{l'\in \Psi_{D}}\sum_{i=1}^{K}\left(\frac{d_{l'l'i}^\alpha}{d_{0l'i}^\alpha}\right)^{\!2}\right) 
\\
&\mkern 100mu		
\left.
\!+\!\left(\! 1 \!+\! \frac{1}{\Delta \tau_c}\!\!\sum\limits_{l'\in\Psi_D}\!\sum\limits_{i =1}^K\! \frac{d_{l'l'i}^\alpha}{d_{0l'i}^\alpha} \!+\!\frac{K(1\! -\! \Delta)}{\Delta\tau_c } \! +\! \frac{\sigma^2}{\Delta\rho\tau_c}\!\right)\!\!\left( \!(1\!-\!\Delta)\!\left(\!K\!+\!\!\!\sum_{l'\in\Psi_{D}}\!\sum_{i=1}^{K}\! \frac{d_{l'l'i}^\alpha}{d_{0l'i}^\alpha}\right) \!+\!\frac{\sigma^2}{\rho}\!\right)
\right)	\stepcounter{equation}\tag{\theequation}\label{eq:SINR_ch_inv_SP_UB}
\end{align*}}

The expected value of the first term in {\color{black}${\SINR_{0k}^\textsc{rp}}^{-1}$} (first term in the denominator of \eqref{eq:SINR_ch_inv_RP}) with RP and the expected value of all terms except the last one in {\color{black}${\SINR_{0k}^\textsc{sp}}^{-1}$ and ${\SINR_{0k}^\textsc{sp-ub}}^{-1}$} (in the denominator of \eqref{eq:SINR_ch_inv_SP} {\color{black}and \eqref{eq:SINR_ch_inv_SP_UB} respectively}) with SP, are given by applying \eqref{eq:d_moments_eq_BS_1} in Lemma~\ref{lem:result_stoc_geom} of Appendix~\ref{app:stoc_geometry}. For the second term in {\color{black}${\SINR_{0k}^\textsc{rp}}^{-1}$} with RP and the last term in {\color{black}${\SINR_{0k}^\textsc{sp}}^{-1}$} and {\color{black}${\SINR_{0k}^\textsc{sp-ub}}^{-1}$} with SP we can apply the following result
{\color{black}\begin{align*}\allowdisplaybreaks
&\E\!\left\{\!\left(\!\sG \! + \!\sT\!\! \sum_{l'\in \Psi_{D}}\!\sum_{i=1}^{K}  \frac{d_{l'l'i}^{\alpha}}{d_{0l'i}^{\alpha}} \!  \right)\!\!
\left(\!\sB\!   +\!\sJ\sum_{l'\in \Psi_{D}}\!\sum_{i=1}^{K} \frac{d_{l'l'i}^{\alpha}}{d_{0l'i}^{\alpha}} \!\right)\! \right\} 
%\\&
\!=\!   \sG \sB \! +\! \sG\sJ\sum_{i=1}^{K} \E\!\left\{\!\sum_{l'\in \Psi_{D}} \frac{  d_{l'l'i}^{\alpha} }{   d_{0l'i}^{\alpha}} \!\right\} \!+\! \sB\sT\sum_{i=1}^{K}\!\E\!\left\{\!\sum_{l'\in \Psi_{D}} \frac{  d_{l'l'i}^{\alpha} }{   d_{0l'i}^{\alpha}}\! \right\}
\\
&\mkern 330mu
\!+\!  \sJ\sT\sum_{i=1}^K\sum_{j=1}^{K}\!\E \!\left\{\! \left(\sum_{l'\in\Psi_{D}}\frac{  d_{l'l'i}^{\alpha}}{ d_{0l'i}^{\alpha} }\right)\!\! \left(\sum_{l\in \Psi_{D}}  \frac{  d_{llj}^{\alpha} }{   d_{0lj}^{\alpha}}\right)\!  \right\} 
\\
&= \!   \sG \sB \!
+\! \sG \sJ \frac{2K}{\alpha - 2}\! + \!\sB\sT \frac{2K}{\alpha - 2}
%\\
%&\mkern 20mu
\!+\! \sJ\sT\sum_{i=1}^{K}\sum_{j=1}^{K}\! \left(\!\E\!\left\{\!\sum_{l'\in\Psi_{D}} \sum_{l\in\Psi_{D}\backslash \{l'\}}\!\frac{d_{l'l'i}^{\alpha}}{d_{0l'i}^{\alpha}} \frac{d_{llj}^{\alpha}}{d_{0lj}^{\alpha}}\!\right\}\! +\!\E\!\left\{\!\sum_{l'\in\Psi_{D}} \frac{d_{l'l'i}^{\alpha}}{d_{0l'i}^{\alpha}} \frac{d_{l'l'j}^{\alpha}}{d_{0l'j}^{\alpha}}\!\right\}\! \right)
\\
&\leq    \sG \sB
+ \sG \sJ \frac{2K}{\alpha - 2} + \sB \sT\frac{2K}{\alpha - 2}
+ \sJ\sT K^2 \left(\left(\frac{2}{\alpha - 2}\right)^2 + \frac{1}{\alpha -1} \right)
\stepcounter{equation}\tag{\theequation}\label{eq:der_LSF_LB_int}
\end{align*}}
which follows from Lemma~\ref{lem:result_stoc_geom}, in Appendix~\ref{app:stoc_geometry}. For the case of RP we have
\begin{equation}
\begin{aligned}
\sG = 1 + \frac{\sigma^2}{\rho \tau_p}, 
\mkern 20mu\sT =\frac{1}{\tau_p},
\mkern 20mu \sB = K+\frac{\sigma^2}{\rho},
\mkern 20mu \sJ= 1,
%\\
\end{aligned}
\label{eq:var_LSF_LB_RP}
\end{equation}
and by combining \eqref{eq:der_LSF_LB_int} with \eqref{eq:var_LSF_LB_RP} we obtain the last two terms in the denominator of ${\underline{\SINR}^\textsc{rp}}$ in \eqref{eq:SINR_LB_rate_lsf_rp}. In the case of SP we have
\begin{equation}
\begin{aligned}
\sG = 1 +\frac{K(1 - \Delta)}{\Delta\tau_c } \! +\! \frac{\sigma^2}{\Delta\rho\tau_c},
\mkern 15mu\sT = \frac{1}{\Delta \tau_c},
\mkern 15mu \sB = K+\frac{\sigma^2}{\rho},
\mkern 15mu \sJ= 1,
\end{aligned}
\label{eq:var_LSF_LB_SP}
\end{equation}
and by combining \eqref{eq:der_LSF_LB_int} with \eqref{eq:var_LSF_LB_SP} the last two terms in the denominators of {\color{black}${\underline{\SINR}^\textsc{sp}}$ in \eqref{eq:SINR_LB_rate_lsf_sp} are obtained. For the case of SP with perfect pilot subtraction we have the same values as in \eqref{eq:var_LSF_LB_SP} except for 
$\sB = K(1 - \Delta)+\frac{\sigma^2}{\rho}$ and $\sJ= 1 - \Delta$. Then by combining these values with \eqref{eq:der_LSF_LB_int} the last two terms in the denominator of ${\underline{\SINR}^\textsc{sp-ub}}$ in \eqref{eq:SINR_LB_rate_lsf_sp_UB} are found.} Thus, the proof is concluded.

\section{Results from Stochastic Geometry}
\label{app:stoc_geometry}
The following results from stochastic geometry are useful.
\begin{lemma}	
	The distribution of the distance $d_{l'l'i}$ between a UEs and its serving BS for $l'\in\Phi_{D}$ and $i \in\{1,\ldots,K\}$, where $\Phi_{D}$ is a homogeneous PPP with density $D$, is $d_{l'l'i}\sim \text{Rayleigh}\left(\frac{1}{\sqrt{2 \pi D}}\right)$. Then we have that $\E\left\{ d_{00k}^{\alpha}\right\} = \frac{\Gamma\left( \alpha/2 + 1\right) }{\left(\pi D\right)^{\alpha/2}}$ for $\alpha > -2$. This result is also found in \cite{Emil_MAMIMO_small_cells}.	  
	\label{lem:exp_d_alpha}
\end{lemma}
\begin{lemma}
	For $\kappa\in \{1,2\}$ and $d_{ll'i}\in \Psi_{D}$ being the distance between $\BS_l$ and $\UE_{l'i}$ where $\Psi_{D}$ describes a the set of BSs distributed as a homogeneous PPP with density $D$, we have	
	\begin{align}
	\E\left\{\sum_{l'\in\Psi_{D}} \left( \frac{d_{l'l'i}^{\alpha}}{ d_{0l'i}^{\alpha}} \right)^{\kappa} \right\} 
	&= \frac{2}{\kappa \alpha - 2}\mkern 56 mu\forall i \in\{1,\ldots,K\}
		\label{eq:d_moments_eq_BS_1}
	\\
	\E \left\{  \sum_{l'\in\Psi_{D}}    \sum_{l\in\Psi_{D}\backslash \{l'\}}  \frac{d_{l'l'i}^{\alpha}}{ d_{0l'i}^{\alpha}} \frac{d_{llj}^{\alpha}}{ d_{0lj}^{\alpha}} \right\} 	&=  \left(\frac{2}{\alpha - 2}\right)^2\mkern 30 mu\forall i,\:j \in\{1,\ldots,K\}
	\label{eq:d_moments_neq_BS_2}
	\\
	\E\left\{ \sum_{l'\in\Psi_{D}} \frac{d_{l'l'i}^{\alpha}}{ d_{0l'i}^{\alpha}} \frac{d_{l'l'j}^{\alpha}}{ d_{0l'j}^{\alpha}} \right\} 
	&\leq \left( \frac{1}{\alpha -  1}\right) \mkern 36 mu\forall i,\:j \in\{1,\ldots,K\} \text{ and } i \neq j\;.
	\label{eq:d_moments_eq_BS_3}
	\end{align}
	\label{lem:result_stoc_geom}		
\end{lemma}
\begin{IEEEproof}
It is found in \cite[Appendix B]{Emil_MAMIMO_small_cells}.
\end{IEEEproof}


\bibliographystyle{IEEEtran}
\bibliography{References}

\begin{thebibliography}{10}
\providecommand{\url}[1]{#1}
\csname url@samestyle\endcsname
\providecommand{\newblock}{\relax}
\providecommand{\bibinfo}[2]{#2}
\providecommand{\BIBentrySTDinterwordspacing}{\spaceskip=0pt\relax}
\providecommand{\BIBentryALTinterwordstretchfactor}{4}
\providecommand{\BIBentryALTinterwordspacing}{\spaceskip=\fontdimen2\font plus
\BIBentryALTinterwordstretchfactor\fontdimen3\font minus
  \fontdimen4\font\relax}
\providecommand{\BIBforeignlanguage}[2]{{%
\expandafter\ifx\csname l@#1\endcsname\relax
\typeout{** WARNING: IEEEtran.bst: No hyphenation pattern has been}%
\typeout{** loaded for the language `#1'. Using the pattern for}%
\typeout{** the default language instead.}%
\else
\language=\csname l@#1\endcsname
\fi
#2}}
\providecommand{\BIBdecl}{\relax}
\BIBdecl

\bibitem{DV_EB_LS_SE_SP_conf}
D.~Verenzuela, E.~Bj\"{o}rnson, and L.~Sanguinetti, ``Spectral efficiency of
  superimposed pilots in uplink massive mimo systems,'' in \emph{Accepted to
  Proc. IEEE GLOBECOM}, 2017.

\bibitem{cisco_GMDT_2014}
Cisco, ``Visual networking index: Global mobile data traffic forecast update,
  2014–2019,'' Tech. Rep., Feb. 2015.

\bibitem{METIS_D8_4}
D.~Aziz, K.~Kusume, O.~Queseth, and et~la., ``D8.4: Metis final project
  report,'' ICT-317669-METIS, Tech. Rep., 2015.

\bibitem{Co2_footprint}
J.~M. A.~Fehske, G.~Fettweis and G.Biczok, ``The global footprint of mobile
  communications: The ecological and economic perspective,'' \emph{IEEE Trans.
  Commun.}, vol.~49, no.~8, pp. 55--62, 2011.

\bibitem{ev_wireless_comm}
R.~Baldemair, E.~Dahlman, G.~Fodor, G.~Mildh, S.~Parkvall, Y.~Selen,
  H.~Tullberg, and K.~Balachandran, ``Evolving wireless communications:
  Addressing the challenges and expectations of the future,'' \emph{IEEE Veh.
  Technol. Mag.}, vol.~8, no.~1, pp. 24--30, March 2013.

\bibitem{Hien_bounds}
H.~Q. Ngo, E.~G. Larsson, and T.~L. Marzetta, ``Energy and spectral efficiency
  of very large multiuser {MIMO} systems,'' \emph{IEEE Trans. Commun.},
  vol.~61, no.~4, pp. 1436--1449, Apr. 2013.

\bibitem{Emil_power_model}
E.~Bj\"{o}rnson, L.~Sanguinetti, J.~Hoydis, and M.~Debbah, ``Optimal design of
  energy-efficient multi-user {MIMO} systems: Is massive {MIMO} the answer?''
  \emph{IEEE Trans. Wireless Commun.}, vol.~14, no.~6, pp. 3059--3075, Jun.
  2015.

\bibitem{Emil_MAMIMO_small_cells}
E.~Bj\"{o}rnson, L.~Sanguinetti, and M.~Kountouris, ``Deploying dense networks
  for maximal energy efficiency: Small cells meet massive {MIMO},'' \emph{IEEE
  J. Sel. Areas Commun.}, vol.~34, no.~4, pp. 832--847, Apr. 2016.

\bibitem{T_marzetta_total_EE}
H.~Yang and T.~L. Marzetta, ``Total energy efficiency of cellular large scale
  antenna system multiple access mobile networks,'' in \emph{Proc. IEEE
  OnlineGreenComm}, Oct 2013, pp. 27--32.

\bibitem{marzetta_MAMIMO}
T.~L. Marzetta, ``Noncooperative cellular wireless with unlimited numbers of
  base station antennas,'' \emph{IEEE Trans. Wireless Commun.}, vol.~9, no.~11,
  pp. 3590--3600, 2010.

\bibitem{T_Marzetta_Ma_MIMO_book}
T.~L. Marzetta, E.~G. Larsson, H.~Yang, and H.~Q. Ngo, \emph{Fundamentals of
  {M}assive {MIMO}}.\hskip 1em plus 0.5em minus 0.4em\relax Cambridge Press,
  2016.

\bibitem{MAMI_low_nr_ant}
H.~Huh, G.~Caire, H.~C. Papadopoulos, and S.~A. Ramprashad, ``Achieving
  "massive {MIMO}" spectral efficiency with a not-so-large number of
  antennas,'' \emph{IEEE Trans. Wireless Commun.}, vol.~11, no.~9, pp.
  3226--3239, September 2012.

\bibitem{coord_ch_est}
H.~Yin, D.~Gesbert, M.~Filippou, and Y.~Liu, ``A coordinated approach to
  channel estimation in large-scale multiple-antenna systems,'' \emph{IEEE J.
  Sel. Areas Commun.}, vol.~31, no.~2, pp. 264--273, February 2013.

\bibitem{Bjornson_pilot_cont_not_lim_jrnl}
\BIBentryALTinterwordspacing
E.~Bj{\"{o}}rnson, J.~Hoydis, and L.~Sanguinetti, ``Massive {MIMO} has
  unlimited capacity,'' \emph{Submitted to IEEE Trans. Wireless Commun.}, 2017.
  [Online]. Available: \url{http://arxiv.org/abs/1705.00538}
\BIBentrySTDinterwordspacing

\bibitem{Julia_pilot_decont}
J.~Vinogradova, E.~Bj\"{o}rnson, and E.~G. Larsson, ``On the separability of
  signal and interference-plus-noise subspaces in blind pilot
  decontamination,'' in \emph{Proc. IEEE ICASSP}, March 2016, pp. 3421--3425.

\bibitem{Hien_EVD_pilot}
H.~Q. Ngo and E.~G. Larsson, ``{EVD}-based channel estimation in multicell
  multiuser {MIMO} systems with very large antenna arrays,'' in \emph{Proc.
  IEEE ICASSP}, March 2012, pp. 3249--3252.

\bibitem{blind_pilot_decont}
R.~R. M\"{u}ller, L.~Cottatellucci, and M.~Vehkaper\"{a}, ``Blind pilot
  decontamination,'' \emph{IEEE J. Sel. Topics Signal Process.}, vol.~8, no.~5,
  pp. 773--786, Oct 2014.

\bibitem{T_marzetta_non_asym}
Y.~Li, Y.-H. Nam, B.~L. Ng, and J.~Zhang, ``A non-asymptotic throughput for
  massive {MIMO} cellular uplink with pilot reuse,'' in \emph{Proc. IEEE
  GLOBECOM}, Dec 2012, pp. 4500--4504.

\bibitem{Emil_pilot_SE}
E.~Bj\"{o}rnson, E.~Larsson, and M.~Debbah, ``Massive {MIMO} for maximal
  spectral efficiency: How many users and pilots should be allocated?''
  \emph{IEEE Trans. Wireless Commun.}, vol.~15, no.~2, pp. 1293--1308, 2016.

\bibitem{Emil_pilot_cluster}
R.~Mochaourab, E.~Bj{\"{o}}rnson, and M.~Bengtsson, ``Adaptive pilot clustering
  in heterogeneous massive {MIMO} networks,'' \emph{IEEE Trans. Wireless
  Commun.}, vol.~15, no.~8, pp. 5555--5568, Aug 2016.

\bibitem{Hoeher99_ch_est_SP}
P.~Hoeher and F.~Tufvesson, ``Channel estimation with superimposed pilot
  sequence,'' in \emph{IEEE Proc. GLOBECOM}, 1999, pp. 2162--2166.

\bibitem{GTZhou03_SP_ch_est_1ord_st}
G.~T. Zhou, M.~Viberg, and T.~McKelvey, ``A first-order statistical method for
  channel estimation,'' \emph{IEEE Signal Process. Lett.}, vol.~10, no.~3, pp.
  57--60, March 2003.

\bibitem{Dong_SiP_SigProcess_04}
M.~Dong, L.~Tong, and B.~M. Sadler, ``Optimal insertion of pilot symbols for
  transmissions over time-varying flat fading channels,'' \emph{IEEE Trans.
  Signal Process.}, vol.~52, no.~5, pp. 1403--1418, May 2004.

\bibitem{SP_stat_fading_MIMO_2017}
A.~T. Asyhari and S.~ten Brink, ``Orthogonal or superimposed pilots? a
  rate-efficient channel estimation strategy for stationary {MIMO} fading
  channels,'' \emph{IEEE Trans. Wireless Commun.}, vol.~16, no.~5, pp.
  2776--2789, May 2017.

\bibitem{SIP_part1_KU_SA}
K.~Upadhya, S.~A. Vorobyov, and M.~Vehkaper\"a, ``Superimposed pilots are
  superior for mitigating pilot contamination in massive {MIMO},'' \emph{IEEE
  Trans. Signal Process.}, vol.~65, no.~11, pp. 2917--2932, June 2017.

\bibitem{SIP_part2_KU_SA}
\BIBentryALTinterwordspacing
K.~Upadhya, S.~A. Vorobyov, and M.~Vehkaper{\"{a}}, ``Downlink performance of
  superimposed pilots in massive {MIMO} systems,'' \emph{Submitted to IEEE
  Trans. Wireless Commun.}, 2016. [Online]. Available:
  \url{http://arxiv.org/abs/1606.04476}
\BIBentrySTDinterwordspacing

\bibitem{VT_SP_approx_2016}
H.~Zhang, S.~Gao, D.~Li, H.~Chen, and L.~Yang, ``On superimposed pilot for
  channel estimation in multicell multiuser {MIMO} uplink: Large system
  analysis,'' \emph{IEEE Trans. Veh. Technol.}, vol.~65, no.~3, pp. 1492--1505,
  March 2016.

\bibitem{FTufvesson15_measured_MAMIMO}
X.~Gao, O.~Edfors, F.~Rusek, and F.~Tufvesson, ``Massive {MIMO} performance
  evaluation based on measured propagation data,'' \emph{IEEE Trans. Wireless
  Commun.}, vol.~14, no.~7, pp. 3899--3911, July 2015.

\bibitem{steve_M_kay}
S.~M. Kay, \emph{Fundamentals of Statistical Signal Processing: Estimation
  Theory}.\hskip 1em plus 0.5em minus 0.4em\relax Prentice Hall, 1993.

\bibitem{Lambert_W}
R.~M. Corless, G.~H. Gonnet, D.~E.~G. Hare, D.~J. Jeffrey, and D.~E. Knuth,
  ``On the {L}ambert{W} function,'' \emph{Adv. in Comput. Math.}, vol.~5,
  no.~1, pp. 329--359, 1996.

\bibitem{M_di_renzo_Exp_stoc_geometry_conf}
W.~Lu and M.~Di~Renzo, ``Stochastic geometry modeling of cellular networks:
  Analysis, simulation and experimental validation,'' in \emph{Proc. ACM
  MSWiM}, 2015, pp. 179--188.

\bibitem{Durisi_MAMIMO_low_res_ADC}
\BIBentryALTinterwordspacing
S.~Jacobsson, G.~Durisi, M.~Coldrey, U.~Gustavsson, and C.~Studer, ``Throughput
  analysis of massive {MIMO} uplink with low-resolution {ADC}s,'' \emph{CoRR},
  vol. abs/1602.01139, 2016. [Online]. Available:
  \url{http://arxiv.org/abs/1602.01139}
\BIBentrySTDinterwordspacing

\bibitem{C_Mollen_1bit_ADC}
C.~Mollén, J.~Choi, E.~G. Larsson, and R.~W. Heath, ``Uplink performance of
  wideband massive {MIMO} with one-bit {ADC}s,'' \emph{IEEE Trans. Wireless
  Commun.}, vol.~16, no.~1, pp. 87--100, Jan 2017.

\end{thebibliography}
\end{document}